\documentclass[useAMS,usegraphicx,usenatbib]{mn2e}

\title[SN~2005bl]{The underluminous Type Ia Supernova 2005bl and the class of objects similar to SN~1991bg\thanks{Based on observations at ESO\,--Paranal, Prog. 075.D-0662(B)}}
\author[Taubenberger et al.]{S. Taubenberger$^{1}$
\thanks{E-mail: tauben@mpa-garching.mpg.de},
S. Hachinger$^{1}$, G. Pignata$^{2,3}$,  P. A. Mazzali$^{1,4}$, C. Contreras$^{5}$, 
\newauthor S. Valenti$^{6,7}$, A. Pastorello$^{1,8}$, N. Elias-Rosa$^{1,9,10}$, O. B\"{a}rnbantner$^{11}$, H. Barwig$^{11}$,  
\newauthor S. Benetti$^{9}$, M. Dolci$^{12}$, J. Fliri$^{11}$, G. Folatelli$^{5}$, W. L. Freedman$^{13}$, S. Gonzalez$^{5}$, 
\newauthor M. Hamuy$^{2}$, W. Krzeminski$^{5}$, N. Morrell$^{5}$, H. Navasardyan$^{9}$, S. E. Persson$^{13}$, 
\newauthor M. M. Phillips$^{5}$, C. Ries$^{11}$, M. Roth$^{5}$, N. B. Suntzeff$^{14}$, M. Turatto$^{9}$ and
\newauthor W. Hillebrandt$^{1}$\\
$^{1}$Max-Planck-Institut f\"{u}r Astrophysik, Karl-Schwarzschild-Str. 1, 85741 Garching bei M\"{u}nchen, Germany\\
$^{2}$Departamento de Astronom\'ia, Universidad de Chile, Casilla 36-D, Santiago, Chile\\
$^{3}$Departamento de Astronom\'ia y Astrof\'isica, Pontificia Universidad Cat\'olica de Chile, Casilla 306, Santiago 22, Chile\\
$^{4}$INAF Osservatorio Astronomico di Trieste, Via Tiepolo 11, 34131 Trieste, Italy\\
$^{5}$Las Campanas Observatory, Carnegie Observatories, Casilla 601, La Serena, Chile\\
$^{6}$European Southern Observatory (ESO), Karl-Schwarzschild-Str. 2, 85748 Garching bei M\"{u}nchen, Germany\\
$^{7}$Physics Department, University of Ferrara, 44100 Ferrara, Italy\\
$^{8}$Astrophysics Research Centre, School of Mathematics and Physics, Queen's University Belfast, Belfast BT7 1NN, UK\\
$^{9}$INAF Osservatorio Astronomico di Padova, Vicolo dell'Osservatorio 5, 35122 Padova, Italy\\
$^{10}$Universidad de La Laguna, Av. Astrof\'isico Francisco S\'anchez s/n, E-38206 La Laguna, Tenerife, Spain\\
$^{11}$Universit\"{a}ts-Sternwarte M\"{u}nchen, Scheinerstr. 1, 81679 M\"{u}nchen, Germany\\
$^{12}$INAF Osservatorio Astronomico di Collurania Teramo, Via Maggini, 64100 Teramo, Italy\\
$^{13}$Observatories of the Carnegie Institution of Washington, Pasadena, CA, USA\\
$^{14}$Texas A\&M University Physics Department, College Station, TX, USA}
\begin{document}

\date{Accepted 2007 November 28; received 2007 November 27; in original form 2007 October 11.}

\pagerange{\pageref{firstpage}--\pageref{lastpage}} \pubyear{2007}

\maketitle

\label{firstpage}

\begin{abstract}
Optical observations of the Type Ia supernova (SN~Ia) 2005bl in 
NGC~4070, obtained from $-6$ to $+66$\,d with respect to the $B$-band 
maximum, are presented. The photometric evolution is characterised 
by rapidly-declining light curves ($\Delta$m$_{15}(B)_\mathrm{true}=
1.93$) and red colours at peak and soon thereafter. With 
M$_{B,\mathrm{max}}=-17.24$ the SN is an underluminous SN~Ia, similar 
to the peculiar SNe~1991bg and 1999by. This similarity also holds for 
the spectroscopic appearance, the only remarkable difference being the 
likely presence of carbon in pre-maximum spectra of SN~2005bl. A 
comparison study among underluminous SNe~Ia is performed, based on a 
number of spectrophotometric parameters. Previously reported 
correlations of the light-curve decline rate with peak luminosity and 
$\mathcal{R}$(Si) are confirmed, and a large range of post-maximum 
Si\,{\sc ii} $\lambda6355$ velocity gradients is encountered. 1D 
synthetic spectra for SN~2005bl are presented, which confirm the 
presence of carbon and suggest an overall low burning efficiency with 
a significant amount of leftover unburned material. Also, the Fe 
content in pre-maximum spectra is very low, which may point to a low 
metallicity of the precursor. Implications for possible progenitor 
scenarios of underluminous SNe~Ia are briefly discussed.
\end{abstract}


\section{Introduction}
\label{Introduction}

The history of underluminous SNe~Ia is a typical example of the ever-recurring 
pattern in which knowledge about nature is accumulated. Usually, in the beginning  
there is the observation of a phenomenon, followed by a successful theoretical 
explanation. However, as further experiments or observations are carried out in 
order to confirm the newly developed theoretical ideas, often an ever higher 
degree of diversity and ever more exceptions from the simple rules are found 
the closer the subject of interest is studied. The need for refined and more 
complex theories to obtain a realistic description of the involved processes 
becomes evident.

In the case of SNe~Ia, first a class of cosmic explosions apparently similar 
in absolute luminosity (``standard candles'') and spectroscopic appearance was 
identified. These events were explained as the disruptions of white dwarfs 
which had accreted matter until they reached their stability limit close to the 
Chandrasekhar mass ($M_\mathrm{Ch}$). However, in 1991 the paradigm of SN~Ia 
homogeneity had to be relaxed a lot. This was triggered by the observation of 
two peculiar SNe~Ia, which thereafter served as prototypes of newly-defined 
SN~Ia subclasses with distinct spectrophotometric properties. One of these, 
SN~1991T (\citealt{Filippenko92a}; \citealt{Phillips92}; \citealt{RuizLapuente92}; 
\citealt*{Mazzali95}), was up to $0.6$ mag brighter than average SNe~Ia, and 
characterised by a hot early-time spectrum with strong Fe\,{\sc iii} features 
and weak or absent Si\,{\sc ii} and S\,{\sc ii} lines. The other one, SN~1991bg 
\citep{Filippenko92b,Leibundgut93,RuizLapuente93,Turatto96,Mazzali97} was even 
more deviant, with low ejecta velocities and a cool spectrum dominated by 
intermediate-mass-element (IME) lines and particularly strong O\,{\sc i} and 
Ti\,{\sc ii}. Moreover, it had unusually red colours at early phases, and was 
underluminous by about $2$ mag at peak (hereafter we will refer to such 
appearance as 91bg-like). Hence, quasi instantaneously the luminosity range of 
SNe~Ia had increased to a factor of ten between the brightest and the faintest 
objects, proving that they were \textit{no} standard candles. However, two years 
later \citet{Phillips93} realised a tight correlation between peak luminosity 
and decline rate in the $B$ band. This relation and revised versions of it 
\citep[e.g.][]{Phillips99} turned SNe~Ia into standardisable candles, and hence 
made them an extremely useful tool for precision cosmology\footnote{Several 
years later even more peculiar objects such as SN~2002cx \citep{Li03} were 
discovered which did not obey the Phillips relation, showing that not \textit{all} 
SNe~Ia are standardisable.}.

In the following years, several more 91bg-like SNe~Ia were discovered, but the 
available data set grew much less rapidly than for ordinary SNe~Ia. From the 
results of the Lick Observatory Supernova Search (LOSS) and the Beijing 
Astronomical Observatory Supernova Survey (BAOSS), \citet{Li01} estimated that 
about $16$\,\% of all SNe~Ia are of the 91bg-like variety. This may still be 
an underestimate, as their low intrinsic luminosity makes 91bg-like SNe prone to 
Malmquist bias; nevertheless \citet{Li01} estimated this effect to be negligible 
in their sample. Statistical studies \citep{Hamuy96a,Hamuy00,Howell01a} have shown 
that SNe~Ia occur in all host-galaxy types, but revealed a correlation between 
SN decline rate and host morphology, with a clear tendency for 91bg-like SNe to 
be associated with early-type hosts and hence old stellar populations.

While the single-degenerate (SD) Chandrasekhar-mass model has survived as the 
favoured scenario for the normal and indeed rather homogeneous SNe~Ia, a number 
of alternative models have been suggested for the 91bg-like subclass. Ideas 
include double-degenerate (DD) explosions of merging white dwarfs, 
sub-Chandrasekhar-mass explosions triggered by detonation of the accreted helium 
layer \citep[cf.][for a review]{Hillebrandt00}, and deflagrations in strongly 
rotating white dwarfs, where the turbulent propagation of the flame front is 
suppressed by the differential rotation \citep{Pfannes06}. Still, the notion 
that 91bg-like SNe are -- in terms of the underlying explosion model -- no 
different from ordinary SNe~Ia, and that the only discriminating parameter is 
the mass of synthesised $^{56}$Ni, has supporters in the SN~Ia community. No 
conclusive evidence for any of these ideas has been found so far.

In this paper we present the joint data set of SN~2005bl obtained by the 
European Supernova Collaboration (ESC)\footnote{http:/$\!$/www.mpa-garching.mpg.de/$\sim$rtn/} 
and the Carnegie Supernova Project (CSP)\footnote{http:/$\!$/www.csp1.lco.cl/$\sim$cspuser1/PUB/CSP.html}. 
Since these observations are among the earliest ever obtained for a 91bg-like 
SN, they may help to better constrain possible progenitor and explosion models. 
The observations and techniques applied for data reduction and calibration are 
discussed in Section~\ref{Observations and data reduction}. In 
Section~\ref{Distance and extinction} we estimate the distance of SN~2005bl and 
the extinction along the line of sight. Sections \ref{Photometric evolution} 
and \ref{Spectroscopic evolution} are devoted to the analysis of the light curves 
and spectra, respectively. Results of 1D spectrum synthesis calculations are 
presented in Section~\ref{Spectral modelling}, and a comparison with other 
underluminous SNe~Ia is performed in Section~\ref{Discussion}, where we also 
discuss the impact of SN~2005bl on our picture of SN~Ia explosions. A short 
summary of the main results is given in Section \ref{Conclusions}.

\section{Observations and data reduction}
\label{Observations and data reduction}

\begin{figure}
   \centering
   \includegraphics[width=8.4cm]{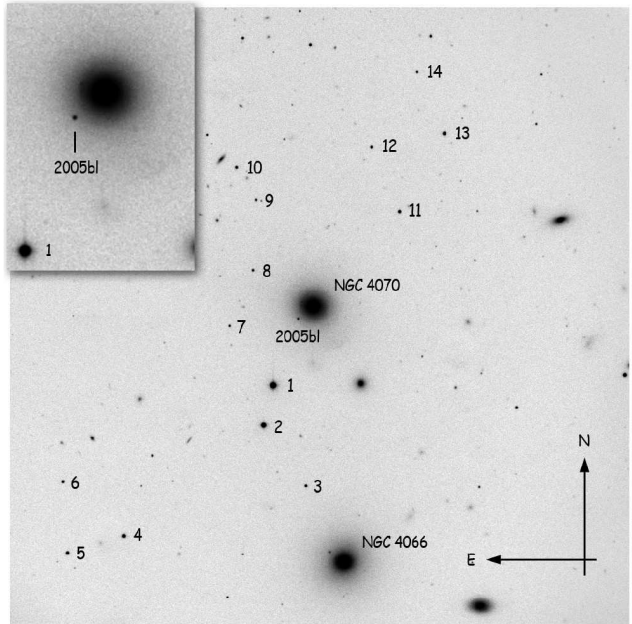}
   \caption{$R$-band image of the SN~2005bl field taken with the Calar Alto 
   2.2\,m Telescope + CAFOS on UT 2005 May 14. The field of view is 9 
   $\times$ 9 arcmin$^2$, and the local sequence stars are indicated. In the 
   upper-left corner a twice-enlarged blow-up of the SN and its host galaxy 
   is shown.}
   \label{fig:chart}
\end{figure}

SN~2005bl ($z$ = 0.024) was discovered in the course of the Lick Observatory
Supernova Search programme (LOSS) with the Katzman Automatic Imaging Telescope 
(KAIT) on UT 2005 April 14.34 and 15.36 at unfiltered magnitudes of $18.8$ and 
$18.3$, respectively \citep{IAUC8512}. The SN was not detected on images 
obtained with the same setup on UT 2005 March $11.33$ to a limiting magnitude 
of $19.5$. Based on spectra taken with the Las Campanas $2.5$\,m du Pont 
Telescope (+\,WFCCD spectrograph) and the Fred Lawrence Whipple Observatory 
$1.5$\,m Telescope (+\,FAST), SN~2005bl was classified as SN Ia, probably 
belonging to the 91bg-like variety given the similarity of the spectra 
with those of SN~1999by a few days before maximum light \citep{IAUC8514_2,
IAUC8514_3}. The SN is located in the elliptical galaxy NGC 4070 (de Vaucouleurs 
morphological type $-4.9$; LEDA\footnote{Lyon-Meudon Extragalactic Database\\ 
\hspace*{0.18cm} http:/$\!$/leda.univ-lyon1.fr/}), projected on a region of 
steep yet smooth background variation (Fig.~\ref{fig:chart}). Details 
on the SN and host galaxy properties are summarised in Table~\ref{properties}.

\begin{table}
\caption{Properties of SN~2005bl and its host galaxy.}
\label{properties}
\begin{center}
\begin{footnotesize}
\begin{tabular}{lrr}
\hline
NGC\,4059\,/\,NGC\,4070$^a$ & & \\
\hline
$\alpha$                              &  12h04m11\fs43                   & 1 \\
$\delta$                              &  +20\degr24\arcmin37\farcs7      & 1 \\
redshift                              &  $0.02406 \pm 0.00008$           & 2 \\
recession velocity $v$                &  $7213 \pm 24$ km\,s$^{-1}$      & 2 \\
$v_\mathrm{Virgo}^b$                  &  $7330 \pm 28$ km\,s$^{-1}$      & 2 \\
$v_\mathrm{CMB}^c$                    &  $7534 \pm 33$ km\,s$^{-1}$      & 2 \\
distance modulus $\mu^d$              &  $35.10 \pm 0.09$ mag            & 2 \\
apparent corr. $B$ magnitude          &  $13.75 \pm 0.15$                & 1 \\
morphological type$^e$                &  E, $-4.9$                       & 1 \\
Galactic reddening $E(B\!-\!V)\!\!\!$ &  $0.028$ mag                     & 3 \\
\hline 
SN~2005bl & & \\
\hline
$\alpha$                              &  12h04m12\fs32                   & 4 \\
$\delta$                              &  +20\degr24\arcmin24\farcs8      & 4 \\
offset from galaxy centre             &  13\farcs9 E, 11\farcs2 S        & 4 \\
host reddening $E(B\!-\!V)$           &  $0.17 \pm 0.08$ mag             & 5 \\
$\Delta$m$_{15}(B)_\mathrm{true}$     &  $1.93 \pm 0.10$                 & 5 \\
JD$_\mathrm{max}$ in $U$              &  $2\,453\,481.4 \pm 0.3$         & 5 \\
JD$_\mathrm{max}$ in $B$              &  $2\,453\,482.6 \pm 0.3$         & 5 \\
JD$_\mathrm{max}$ in $V$              &  $2\,453\,484.9 \pm 0.3$         & 5 \\
JD$_\mathrm{max}$ in $R$              &  $2\,453\,485.9 \pm 0.3$         & 5 \\
JD$_\mathrm{max}$ in $I$              &  $2\,453\,487.0 \pm 0.3$         & 5 \\
JD$_\mathrm{max}$ in $g$              &  $2\,453\,483.0 \pm 0.3$         & 5 \\
JD$_\mathrm{max}$ in $z$              &  $2\,453\,487.0 \pm 3.0$         & 5 \\
$U_\mathrm{max}$                      &  $19.14 \pm 0.18$                & 5 \\ 
$B_\mathrm{max}$                      &  $18.68 \pm 0.04$                & 5 \\ 
$V_\mathrm{max}$                      &  $17.87 \pm 0.03$                & 5 \\ 
$R_\mathrm{max}$                      &  $17.55 \pm 0.03$                & 5 \\ 
$I_\mathrm{max}$                      &  $17.38 \pm 0.04$                & 5 \\ 
$g_\mathrm{max}$                      &  $18.20 \pm 0.07$                & 5 \\ 
$z_\mathrm{max}$                      &  $17.77 \pm 0.12$                & 5 \\ 
M$_{U,\mathrm{max}}$                  &  $-16.91 \pm 0.43$               & 5 \\
M$_{B,\mathrm{max}}$                  &  $-17.24 \pm 0.34$               & 5 \\
M$_{V,\mathrm{max}}$                  &  $-17.85 \pm 0.27$               & 5 \\
M$_{R,\mathrm{max}}$                  &  $-18.06 \pm 0.23$               & 5 \\
M$_{I,\mathrm{max}}$                  &  $-18.10 \pm 0.18$               & 5 \\
M$_{g,\mathrm{max}}$                  &  $-17.55 \pm 0.29$               & 5 \\
M$_{z,\mathrm{max}}$                  &  $-17.67 \pm 0.20$               & 5 \\
\hline  
\end{tabular}
\\[1.5ex]
$^a$ The galaxy is listed twice in the NGC catalogue.\\
$^b$ $v$ corrected for Local-Group infall onto Virgo cluster\\
$^c$ $v$ corrected to the CMB reference frame\\
$^d$ from $v_\mathrm{CMB}$, using $H_0=72\,\rmn{km}\,\rmn{s}^{-1}\rmn{Mpc}^{-1}$\\
$^e$ numerical code according to de Vaucouleurs\\[1.8ex] 
1: LEDA; 2: NED\footnotemark; 3: \citealt*{Schlegel98}; 
4: \citealt{IAUC8515}; 5: this work 
\end{footnotesize}
\end{center} 
\end{table}
\footnotetext{NASA/IPAC Extragalactic Database\\ 
\hspace*{0.18cm} http:/$\!$/nedwww.ipac.caltech.edu/}

Although SN~2005bl was too distant to fulfil the formal selection criteria of 
the ESC, optical follow-up observations were performed owing to the peculiarities 
found in our first spectrum, obtained almost at the same time as the classification 
spectra by other groups. However, given the dimness of the SN, an intensive coverage 
as for other ESC targets was out of reach. In particular, the early part of 
the light curves was not well sampled owing to bad weather and scheduling 
constraints. At the same time, the CSP collaboration started their follow-up of 
SN~2005bl, focussing mainly on the photometric evolution near maximum light. 
Hence, the two data sets were almost perfectly complementary.

\subsection{Photometric data}
\label{Photometry}

Optical photometry of SN~2005bl was acquired from one week before to about two 
months after maximum light in $B$. The basic data reduction (bias subtraction, 
overscan correction and flat-fielding) was performed using standard routines in 
{\sc iraf}\footnote{{\sc iraf} is distributed by the National Optical Astronomy 
Observatories, which are operated by the Association of Universities for 
Research in Astronomy, Inc, under contract to the National Science 
Foundation.}\citep{IRAFphot,IRAFccdred}. The local sequence of stars in the SN 
field shown in Fig.~\ref{fig:chart} was calibrated with respect to a number of 
\citet[][for $U\!BV\!RI$]{Landolt92} and Sloan \citep[][for $ugriz$]{Smith02} 
standard fields on several photometric nights. The magnitudes of the calibrated 
local sequence, listed in Table~\ref{sequence}, were used subsequently to 
determine the SN magnitudes in a relative measurement. For the sequence-star 
magnitudes in the Sloan $ugri$ bands, the reader is referred to Contreras et al. 
(in prep.).
\begin{table*}
\caption{Magnitudes of the local sequence stars in the field of
SN~2005bl (Fig.~\ref{fig:chart}).} 
\begin{footnotesize}
\begin{tabular}{ccccccc}
\hline
ID & $U$ & $B$ & $V$ & $R$ & $I$ & $z$\\
\hline
 1  &  15.207 $\pm$ 0.022 & 14.786 $\pm$ 0.013 & 13.952 $\pm$ 0.014 & 13.502 $\pm$ 0.010 & 13.080 $\pm$ 0.013 & 13.386 $\pm$ 0.013\\
 2  &  16.387 $\pm$ 0.022 & 15.931 $\pm$ 0.012 & 15.048 $\pm$ 0.014 & 14.510 $\pm$ 0.011 & 14.011 $\pm$ 0.012 & 14.295 $\pm$ 0.024\\
 3  &  18.340 $\pm$ 0.019 & 18.633 $\pm$ 0.021 & 18.110 $\pm$ 0.016 & 17.767 $\pm$ 0.013 & 17.385 $\pm$ 0.012 & 17.710 $\pm$ 0.074\\
 4  &                     & 18.154 $\pm$ 0.014 & 16.658 $\pm$ 0.020 & 15.665 $\pm$ 0.021 & 14.792 $\pm$ 0.017 &                   \\
 5  &                     & 17.464 $\pm$ 0.009 & 16.962 $\pm$ 0.027 & 16.621 $\pm$ 0.007 & 16.269 $\pm$ 0.012 &                   \\
 6  &                     & 19.132 $\pm$ 0.014 & 18.220 $\pm$ 0.015 & 17.690 $\pm$ 0.018 & 17.247 $\pm$ 0.010 &                   \\
 7  &  19.274 $\pm$ 0.017 & 19.488 $\pm$ 0.020 & 18.979 $\pm$ 0.014 & 18.629 $\pm$ 0.008 & 18.272 $\pm$ 0.015 & 18.526 $\pm$ 0.035\\
 8  &  19.538 $\pm$ 0.022 & 19.396 $\pm$ 0.021 & 18.659 $\pm$ 0.019 & 18.251 $\pm$ 0.013 & 17.824 $\pm$ 0.017 & 18.096 $\pm$ 0.030\\
 9  &  18.656 $\pm$ 0.049 & 18.924 $\pm$ 0.016 & 18.454 $\pm$ 0.018 & 18.115 $\pm$ 0.020 & 17.834 $\pm$ 0.018 & 18.112 $\pm$ 0.040\\
10  &                     & 20.035 $\pm$ 0.020 & 18.640 $\pm$ 0.022 & 17.532 $\pm$ 0.014 & 16.260 $\pm$ 0.022 &                   \\
11  &  18.355 $\pm$ 0.049 & 18.267 $\pm$ 0.013 & 17.539 $\pm$ 0.021 & 17.104 $\pm$ 0.015 & 16.687 $\pm$ 0.011 & 16.965 $\pm$ 0.014\\
12  &                     & 18.346 $\pm$ 0.014 & 17.873 $\pm$ 0.020 & 17.510 $\pm$ 0.016 & 17.196 $\pm$ 0.011 &                   \\
13  &                     & 18.278 $\pm$ 0.011 & 17.206 $\pm$ 0.017 & 16.509 $\pm$ 0.018 & 15.893 $\pm$ 0.025 &                   \\
14  &                     & 19.083 $\pm$ 0.013 & 18.198 $\pm$ 0.023 & 17.666 $\pm$ 0.013 & 17.193 $\pm$ 0.020 &                   \\
\hline\\[-0.7ex]
\end{tabular}
\end{footnotesize}
\label{sequence}
\end{table*}

Although the host-galaxy background at the projected SN site seemed to be 
smooth, we applied the template-subtraction technique \citep{Filippenko86} in 
order to eliminate any possible contamination from the host galaxy, a justified 
concern given the faintness of the SN. For this purpose we acquired templates 
in $BV\!RI$ with CAFOS mounted on the Calar Alto 2.2\,m Telescope on UT 2006 
March 28, and in $uBV\!gri$ with the Las Campanas 2.5\,m du Pont Telescope from 
UT 2006 April 3 to 30, about one year after the explosion when the SN had 
faded from visibility. For the ESC data, the galaxy subtraction was performed 
using the {\sc iraf} plug-in {\sc svsub} written by S.V. (based on {\sc isis}). 
The instrumental SN magnitudes were determined in the background-subtracted 
images with point-spread function (PSF) fitting photometry using the software 
package {\sc snoopy}, specifically designed for this purpose by F. Patat and 
implemented in {\sc iraf} by E. Cappellaro. For the $z$ band no templates were 
available, and the measurements were performed with ordinary background-fitting 
PSF photometry in {\sc snoopy}. A check in the $BV\!RI$ bands showed a good 
agreement between the two methods. Therefore, we are confident that also our 
$z$-band photometry is sufficiently reliable.

\begin{figure}   
   \centering
   \includegraphics[width=8.4cm]{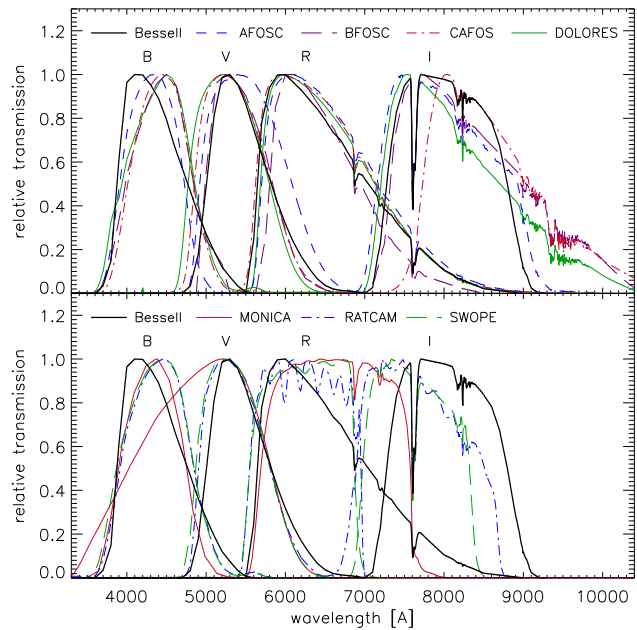}
   \caption{Instrumental $BV\!RI$\,/\,$ri$ passbands used for the observations 
   of SN~2005bl. The standard \citet{Bessell90} curves are also displayed in 
   the figure for comparison.}
   \label{fig:filters}
\end{figure}

The calibration of the SN magnitudes to the desired standard photometric 
systems, the Johnson/Cousins system \citep{Bessell90} for $U\!BV\!RI$ and the 
Sloan system \citep{Sloan96} for $g$ and $z$, was complicated by the variety 
of filters mounted at the various telescopes, some of which deviated strongly 
from the standard. Fig.~\ref{fig:filters} shows the $BV\!RI$ and $ri$ response 
curves of all instruments used for the SN follow-up. 

\subsection*{Bessell photometry}

In order to compensate for the differences in the transmission curves, and to 
report the magnitudes on the Bessell system, we made use of the ``$S$-correction'' 
technique based on the prescription of \citet{Stritzinger02}, \citet{Pignata04b}, 
and references therein. $S$-corrections were computed for the $BV\!RI$ and Sloan 
$ri$ data, based on our spectra of SN~2005bl. Unfortunately, the spectra do not 
fully cover the $U$-band region, so that no $U$-band $S$-correction could be 
derived. Instead, transformation formulae \citep*{Jordi06,Zhao06} were employed to 
convert the Swope $u$-band data to the Bessell system. Similarly, no $S$-correction 
was applied to data later than $+50$\,d, since our spectroscopic follow-up ends 
already well before. The strong deviation of the MONICA $V$ filter (actually 
a Roeser $BV$ filter) from the Bessell description (see Fig.~\ref{fig:filters}), 
along with some lacking information required to reconstruct the response curve, 
makes the MONICA $V$-band correction less reliable. 

The redshift of SN~2005bl is not negligible. Therefore, in computing the 
$S$-correction the spectra were shifted in order to account also for the 
$K$-correction. Since the same restrictions as before apply also here, no 
$K$-correction was performed for the $U$ band and for any data later than $+50$\,d.

In Table~\ref{SN_mags1} the fully calibrated and -- whenever possible -- $S$- 
and $K$-corrected $U\!BV\!RI$ Bessell magnitudes of SN~2005bl are reported, 
together with their uncertainties. Both the intrinsic Bessell data and the 
transformed Sloan photometry entered into this Table. Table~\ref{S-corr} shows 
the combined $S$- and $K$-correction, i.e., the difference between the magnitudes 
of Table~\ref{SN_mags1} and those obtained with a first-order colour-term 
calibration. The differences are significant, so that the lack of the $S$- and 
$K$-correction for $U$ may introduce considerable uncertainties.

\begin{table*}
\caption{$S$- and $K$-corrected Bessell magnitudes of SN~2005bl.$^a$} 
\label{SN_mags1}
\begin{footnotesize}
\begin{tabular}{rrcccccll}
\hline
JD$^b$\ & Epoch$^c$  &         $U$         &        $B$         &        $V$         &        $R$         &        $I$         & Telescope  & Seeing$^d$\\
\hline
476.61\ & $-6.0$\ \ \ & 19.591 $\pm$ 0.114 & 19.364 $\pm$ 0.022 & 18.950 $\pm$ 0.030 & 18.776 $\pm$ 0.024 & 18.599 $\pm$ 0.045 & \ \ SWO    & \ \ 1.41\\
477.63\ & $-5.0$\ \ \ & 19.389 $\pm$ 0.117 & 19.135 $\pm$ 0.022 & 18.695 $\pm$ 0.029 & 18.523 $\pm$ 0.023 & 18.402 $\pm$ 0.038 & \ \ SWO    & \ \ 1.34\\
478.66\ & $-3.9$\ \ \ & 19.237 $\pm$ 0.134 & 18.955 $\pm$ 0.022 & 18.461 $\pm$ 0.029 & 18.290 $\pm$ 0.022 & 18.120 $\pm$ 0.036 & \ \ SWO    & \ \ 1.68\\
479.64\ & $-3.0$\ \ \ &                    & 18.818 $\pm$ 0.029 & 18.307 $\pm$ 0.031 & 18.116 $\pm$ 0.030 & 17.948 $\pm$ 0.043 & \ \ SWO    & \ \ 1.57\\
479.68\ & $-2.9$\ \ \ &                    &                    & 18.393 $\pm$ 0.042 & 18.120 $\pm$ 0.030 & 17.883 $\pm$ 0.034 & \ \ TNG    & \ \ 1.47\\
480.65\ & $-1.9$\ \ \ & 19.151 $\pm$ 0.144 & 18.766 $\pm$ 0.025 & 18.150 $\pm$ 0.029 & 17.883 $\pm$ 0.028 & 17.770 $\pm$ 0.038 & \ \ SWO    & \ \ 1.44\\
481.61\ & $-1.0$\ \ \ & 19.185 $\pm$ 0.174 & 18.709 $\pm$ 0.027 & 18.052 $\pm$ 0.030 & 17.798 $\pm$ 0.022 & 17.650 $\pm$ 0.036 & \ \ SWO    & \ \ 1.45\\
483.62\ &  $1.0$\ \ \ & 19.235 $\pm$ 0.203 & 18.682 $\pm$ 0.040 & 17.899 $\pm$ 0.030 & 17.636 $\pm$ 0.026 & 17.513 $\pm$ 0.039 & \ \ SWO    & \ \ 1.54\\
484.65\ &  $2.1$\ \ \ & 19.296 $\pm$ 0.291 & 18.733 $\pm$ 0.042 & 17.839 $\pm$ 0.031 & 17.526 $\pm$ 0.043 & 17.414 $\pm$ 0.045 & \ \ SWO    & \ \ 1.52\\
485.39\ &  $2.8$\ \ \ &                    & 18.872 $\pm$ 0.036 & 17.887 $\pm$ 0.051 & 17.563 $\pm$ 0.035 & 17.361 $\pm$ 0.051 & \ \ LT     & \ \ 0.84\\
489.32\ &  $6.7$\ \ \ &                    &                    & 18.172 $\pm$ 0.280 &                    &                    & \ \ WD     & \ \ 1.89\\
489.59\ &  $7.0$\ \ \ & 20.248 $\pm$ 0.152 & 19.472 $\pm$ 0.031 & 18.146 $\pm$ 0.027 & 17.692 $\pm$ 0.021 & 17.416 $\pm$ 0.035 & \ \ SWO    & \ \ 1.84\\
490.52\ &  $7.9$\ \ \ & 20.640 $\pm$ 0.155 & 19.653 $\pm$ 0.026 & 18.194 $\pm$ 0.025 & 17.781 $\pm$ 0.025 & 17.435 $\pm$ 0.044 & \ \ LT     & \ \ 0.81\\
490.60\ &  $8.0$\ \ \ & 20.640 $\pm$ 0.163 & 19.687 $\pm$ 0.023 & 18.258 $\pm$ 0.028 & 17.791 $\pm$ 0.022 & 17.456 $\pm$ 0.038 & \ \ SWO    & \ \ 1.34\\
491.51\ &  $8.9$\ \ \ &                    & 19.824 $\pm$ 0.046 & 18.296 $\pm$ 0.048 & 17.845 $\pm$ 0.027 &                    & \ \ WD     & \ \ 1.60\\
491.57\ &  $9.0$\ \ \ & 20.620 $\pm$ 0.140 & 19.834 $\pm$ 0.022 & 18.359 $\pm$ 0.029 & 17.836 $\pm$ 0.026 & 17.477 $\pm$ 0.039 & \ \ SWO    & \ \ 1.37\\
493.35\ & $10.8$\ \ \ &                    &                    & 18.553 $\pm$ 0.028 & 17.959 $\pm$ 0.025 & 17.503 $\pm$ 0.031 & \ \ LOI    & \ \ 2.41\\
498.38\ & $15.8$\ \ \ &                    &                    &                    & 18.551 $\pm$ 0.045 & 17.805 $\pm$ 0.052 & \ \ LT     & \ \ 0.95\\
500.51\ & $17.9$\ \ \ &                    &                    & 19.258 $\pm$ 0.184 & 18.722 $\pm$ 0.041 &                    & \ \ WD     & \ \ 2.38\\
502.38\ & $19.8$\ \ \ &                    &                    &                    & 18.807 $\pm$ 0.034 & 18.192 $\pm$ 0.063 & \ \ LT     & \ \ 0.70\\
503.35\ & $20.8$\ \ \ &                    & 20.912 $\pm$ 0.073 & 19.334 $\pm$ 0.040 & 18.960 $\pm$ 0.023 &                    & \ \ WD     & \ \ 1.55\\
503.35\ & $20.8$\ \ \ &                    & 20.893 $\pm$ 0.133 & 19.429 $\pm$ 0.046 & 18.965 $\pm$ 0.074 & 18.361 $\pm$ 0.059 & \ \ Ekar   & \ \ 2.57\\
504.49\ & $21.9$\ \ \ &                    &                    &                    & 19.035 $\pm$ 0.035 & 18.325 $\pm$ 0.057 & \ \ LT     & \ \ 0.67\\
505.49\ & $22.9$\ \ \ &                    & 20.936 $\pm$ 0.050 & 19.609 $\pm$ 0.069 & 19.180 $\pm$ 0.045 & 18.474 $\pm$ 0.064 & \ \ Caha   & \ \ 1.29\\
506.41\ & $23.8$\ \ \ &                    & 21.052 $\pm$ 0.081 & 19.597 $\pm$ 0.037 & 19.216 $\pm$ 0.028 & 18.571 $\pm$ 0.035 & \ \ Caha   & \ \ 1.91\\
510.53\ & $27.9$\ \ \ &                    &                    & 19.693 $\pm$ 0.119 &                    &                    & \ \ WD     & \ \ 1.70\\
511.36\ & $28.8$\ \ \ &                    &                    & 19.687 $\pm$ 0.092 & 19.530 $\pm$ 0.045 &                    & \ \ WD     & \ \ 1.70\\
512.55\ & $29.9$\ \ \ &                    & 21.168 $\pm$ 0.176 & 19.847 $\pm$ 0.063 & 19.480 $\pm$ 0.036 & 18.874 $\pm$ 0.051 & \ \ SWO    & \ \ 1.56\\
516.48\ & $33.9$\ \ \ &                    &                    &                    & 19.877 $\pm$ 0.073 &                    & \ \ WD     & \ \ 1.78\\
518.43\ & $35.8$\ \ \ &                    &                    & 20.044 $\pm$ 0.068 & 20.020 $\pm$ 0.042 &                    & \ \ WD     & \ \ 1.52\\
519.40\ & $36.8$\ \ \ &                    & 21.421 $\pm$ 0.075 & 20.031 $\pm$ 0.042 & 19.925 $\pm$ 0.055 & 19.310 $\pm$ 0.096 & \ \ LT     & \ \ 1.01\\
520.44\ & $37.8$\ \ \ &                    & 21.407 $\pm$ 0.124 & 20.205 $\pm$ 0.055 & 20.002 $\pm$ 0.045 &                    & \ \ WD     & \ \ 2.07\\
521.39\ & $38.8$\ \ \ &                    &                    & 20.122 $\pm$ 0.040 & 20.009 $\pm$ 0.036 & 19.358 $\pm$ 0.094 & \ \ LT     & \ \ 0.96\\
524.42\ & $41.8$\ \ \ &                    & 21.379 $\pm$ 0.097 & 20.141 $\pm$ 0.069 & 20.159 $\pm$ 0.071 &                    & \ \ WD     & \ \ 1.40\\
538.38\ & $55.8$\ \ \ &                    &                    & 20.685 $\pm$ 0.168 &                    &                    & \ \ WD     & \ \ 1.75\\
548.41\ & $65.8$\ \ \ &                    &                    & 20.984 $\pm$ 0.280 & 21.183 $\pm$ 0.347 &                    & \ \ WD     & \ \ 2.15\\
\hline 
\end{tabular}
\\[1.4ex]
$^a$ No $S$- and $K$-correction applied in the $U$ band and to any data after $+50$\,d.\quad
$^b$ JD $-$ 2\,453\,000.00\quad
$^c$ Epoch in days with respect to the $B$-band maximum JD $2\,453\,482.6 \pm 0.5$.\quad
$^d$ Average seeing in arcsec over all filter bands.\\[1.6ex]
SWO = Las Campanas 1.0\,m Swope Telescope + CCD; \,http:/$\!$/www.lco.cl/telescopes-information/henrietta-swope/\\
TNG = 3.58\,m Telescopio Nazionale Galileo + DOLORES; \,http:/$\!$/www.tng.iac.es/instruments/lrs/\\
LT = 2.0\,m Liverpool Telescope + RATCAM; \,http:/$\!$/telescope.livjm.ac.uk/Info/TelInst/Inst/RATCam/\\
WD = 0.8\,m Wendelstein Telescope + MONICA; \,http:/$\!$/www.wendelstein-observatorium.de/monica/monica\_en.html\\
LOI = 1.52\,m Loiano Telescope + BFOSC; \,http:/$\!$/www.bo.astro.it/loiano/152cm.html\\
Ekar = Asiago 1.82\,m Telescope + AFOSC; \,http:/$\!$/www.oapd.inaf.it/asiago/2000/2300/2310.html\\
Caha = Calar Alto 2.2\,m Telescope + CAFOS SiTe; \,http:/$\!$/www.caha.es/CAHA/Instruments/CAFOS/\\
\end{footnotesize}
\end{table*}

\begin{table}
\caption{Amount of $S$- and $K$-correction contained in the magnitudes reported 
        in Table~\ref{SN_mags1}. No $S$- and $K$-correction has been applied to 
	the $U$-band data.}
\label{S-corr}
\begin{footnotesize}
\begin{tabular}{rrrrrrl}
\hline
JD$^a$ & Epoch$^b \!\!\!$ & $B$\quad\ & $V$\quad\ & $R$\quad\ & $I$\quad\ & Tel.$^c$\\
\hline
476.6 & $-6.0$\ \ & $-0.020$  & $ 0.030$  &  $0.033$  & $ 0.147$  & SWO    \\
477.6 & $-5.0$\ \ & $-0.038$  & $ 0.029$  &  $0.044$  & $ 0.177$  & SWO    \\
478.7 & $-3.9$\ \ & $-0.028$  & $ 0.011$  &  $0.045$  & $ 0.162$  & SWO    \\
479.6 & $-3.0$\ \ & $-0.054$  & $ 0.011$  &  $0.044$  & $ 0.151$  & SWO    \\
479.7 & $-2.9$\ \ &           & $ 0.123$  &  $0.193$  & $ 0.252$  & TNG    \\
480.7 & $-1.9$\ \ & $-0.032$  & $ 0.007$  &  $0.022$  & $ 0.093$  & SWO    \\
481.6 & $-1.0$\ \ & $-0.062$  & $-0.003$  &  $0.043$  & $ 0.110$  & SWO    \\
483.6 &  $1.0$\ \ & $-0.078$  & $-0.044$  &  $0.088$  & $ 0.102$  & SWO    \\
484.7 &  $2.1$\ \ & $-0.099$  & $-0.042$  &  $0.048$  & $ 0.085$  & SWO    \\
485.4 &  $2.8$\ \ & $-0.077$  & $-0.002$  &  $0.136$  & $-0.054$  & LT     \\
489.3 &  $6.7$\ \ &           & $ 0.137$  &           &           & WD     \\
489.6 &  $7.0$\ \ & $-0.113$  & $-0.039$  &  $0.187$  & $ 0.060$  & SWO    \\
490.5 &  $7.9$\ \ & $-0.116$  & $-0.034$  &  $0.166$  & $-0.042$  & LT     \\
490.6 &  $8.0$\ \ & $-0.128$  & $-0.065$  &  $0.213$  & $ 0.049$  & SWO    \\
491.5 &  $8.9$\ \ & $-0.233$  & $-0.034$  &  $0.074$  &           & WD     \\
491.6 &  $9.0$\ \ & $-0.142$  & $-0.074$  &  $0.190$  & $ 0.035$  & SWO    \\
493.4 & $10.8$\ \ &           & $-0.029$  &  $0.027$  & $-0.018$  & LOI    \\
498.4 & $15.8$\ \ &           &           &  $0.126$  & $-0.105$  & LT     \\
500.5 & $17.9$\ \ &           & $-0.064$  &  $0.096$  &           & WD     \\
502.4 & $19.8$\ \ &           &           &  $0.096$  & $-0.209$  & LT     \\
503.4 & $20.8$\ \ & $-0.237$  & $-0.065$  &  $0.074$  &           & WD     \\
503.4 & $20.8$\ \ & $-0.166$  & $-0.067$  &  $0.055$  & $-0.017$  & Ekar   \\
504.5 & $21.9$\ \ &           &           &  $0.113$  & $-0.220$  & LT     \\
505.5 & $22.9$\ \ & $-0.201$  & $-0.087$  &  $0.083$  & $ 0.009$  & Caha   \\
506.4 & $23.8$\ \ & $-0.207$  & $-0.085$  &  $0.093$  & $ 0.011$  & Caha   \\
510.5 & $27.9$\ \ &           & $-0.292$  &           &           & WD     \\
511.4 & $28.8$\ \ &           & $-0.286$  &  $0.062$  &           & WD     \\
512.6 & $29.9$\ \ & $-0.150$  & $-0.084$  &  $0.199$  & $-0.218$  & SWO    \\
516.5 & $33.9$\ \ &           &           &  $0.078$  &           & WD     \\
518.4 & $35.8$\ \ &           & $-0.037$  &  $0.082$  &           & WD     \\
519.4 & $36.8$\ \ & $-0.125$  & $-0.038$  &  $0.120$  & $-0.304$  & LT     \\
520.4 & $37.8$\ \ & $-0.196$  & $-0.076$  &  $0.076$  &           & WD     \\
521.4 & $38.8$\ \ &           & $-0.078$  &  $0.127$  & $-0.304$  & LT     \\
524.4 & $41.8$\ \ & $-0.191$  & $-0.054$  &  $0.074$  &           & WD     \\
\hline 
\end{tabular}
\\[1.5ex]
$^a$ JD $-$ 2\,453\,000.0\quad
$^b$ Epoch in days with respect to the estimated $B$-band maximum JD $2\,453\,482.6 \pm 0.5$.\quad
$^c$ See Table~\ref{SN_mags1} for details.
\end{footnotesize}
\end{table}

\subsection*{Sloan photometry}

Besides transforming the Swope and Liverpool Sloan-filter observations to the 
Bessell system via $S$-corrections, light curves in the Sloan photometric 
system itself were constructed. The magnitudes were calibrated through 
first-order colour equations without employing $S$-corrections, since at both 
telescopes the filters are close to the standard Sloan prescription. 
Table~\ref{SN_mags2} reports the Sloan $griz$ photometry from the Liverpool 
telescope. For the original Swope photometry see Contreras et al. (in prep.).

\begin{table*}
\caption{Sloan photometry of SN~2005bl obtained with the Liverpool telescope. 
        No $S$- and $K$-correction has been applied. For the Swope data see 
	Contreras et~al. (in prep.).} 
\label{SN_mags2}
\begin{footnotesize}
\begin{tabular}{rrcccc}
\hline
JD$^a$\ & Epoch$^b$  &      $g$         &      $r$         &      $i$         &      $z$        \\
\hline
485.4\ &  $2.8$\ \ \ & 18.295 $\pm$ 0.115 & 17.669 $\pm$ 0.023 & 17.866 $\pm$ 0.051 & 17.769 $\pm$ 0.080\\
490.5\ &  $7.9$\ \ \ &                    & 17.877 $\pm$ 0.025 & 17.870 $\pm$ 0.031 & 17.808 $\pm$ 0.039\\
498.4\ & $15.8$\ \ \ &                    & 18.676 $\pm$ 0.046 & 18.308 $\pm$ 0.092 & 18.113 $\pm$ 0.080\\
502.4\ & $19.8$\ \ \ &                    & 18.921 $\pm$ 0.039 & 18.760 $\pm$ 0.059 &                   \\
504.5\ & $21.9$\ \ \ &                    & 19.138 $\pm$ 0.047 & 18.878 $\pm$ 0.077 & 18.548 $\pm$ 0.095\\
519.4\ & $36.8$\ \ \ &                    & 19.998 $\pm$ 0.071 & 19.982 $\pm$ 0.093 & 19.548 $\pm$ 0.165\\
521.4\ & $38.8$\ \ \ &                    & 20.073 $\pm$ 0.043 & 20.081 $\pm$ 0.100 & 19.680 $\pm$ 0.240\\
\hline 
\end{tabular}
\\[1.5ex]
$^a$ JD $-$ 2\,453\,000.0\quad
$^b$ Epoch in days with respect to $B$-band maximum (JD $=2\,453\,482.6 \pm 0.5$).
\end{footnotesize}
\end{table*}

\subsection{Spectroscopic data}
\label{Spectroscopic data}

\begin{table*}
\caption{Journal of spectroscopic observations of SN~2005bl.}
\label{spectra}
\begin{footnotesize}
\begin{tabular}{lrrlcccccl}
\hline 
UT Date   & JD$^a$\ \ \ & Epoch$^b$     & Exposure              & Airmass & Tel.      & Grism          &  Range [\AA]       & Res.\,[\AA]$^c$ & Standards   \\[0.8ex]
\hline
05/04/16  & 476.6\ \ \  &$-6.0$\ \ \ \  & \ \,900\,s $\times$ 3 &  1.54   &  DUP      & blue           &  3800 --  9200     &  6             & L745-46A, LTT7987 \\
05/04/17  & 477.6\ \ \  &$-5.0$\ \ \ \  & 2400\,s               &  1.54   &  Caha     & b200           &  3500 --  8800     & 10             & BD+33\,2642 \\
05/04/19  & 479.6\ \ \  &$-3.0$\ \ \ \  & \ \,900\,s $\times$ 3 &  1.54   &  DUP      & blue           &  3700 --  9000     &  6             & L745-46A, LTT7379 \\
05/04/19  & 479.7\ \ \  &$-2.9$\ \ \ \  & 1200\,s               &  2.06   &  TNG      & LR-R           &  5000 --  9750     & 11             & Hz44        \\
05/04/26  & 487.4\ \ \  &$ 4.8$\ \ \ \  & 1500\,s               &  1.09   &  TNG      & LR-B           &  3300 --  8000     & 13             & Feige\,34   \\
05/04/26  & 487.4\ \ \  &$ 4.8$\ \ \ \  & 1500\,s               &  1.15   &  TNG      & LR-R           &  5000 --  9750     & 12             & Feige\,34   \\
05/05/04  & 495.5\ \ \  &$12.9$\ \ \ \  & 1500\,s               &  1.04   &  TNG      & LR-B           &  3500 --  8900     & 14             & Feige\,56   \\
05/05/04  & 495.5\ \ \  &$12.9$\ \ \ \  & 1500\,s               &  1.08   &  TNG      & LR-R           &  5000 --  9750     & 12             & Feige\,56   \\
05/05/11  & 502.4\ \ \  &$19.8$\ \ \ \  & 1500\,s $\times$ 2    &  1.01   &  TNG      & LR-B           &  3300 --  8000     & 14             & Feige\,66   \\
05/05/11  & 502.4\ \ \  &$19.8$\ \ \ \  & 2700\,s $\times$ 2    &  1.14   &  Caha     & r200           &  6200 --  9750     & 11             & Feige\,34   \\
05/05/14  & 505.4\ \ \  &$22.8$\ \ \ \  & 2400\,s $\times$ 2    &  1.11   &  Caha     & b200           &  3500 --  8800     & 14             & BD+33\,2642 \\
05/05/24  & 515.6\ \ \  &$33.0$\ \ \ \  & 1800\,s               &  1.93   &  VLT      & 300V\,+\,GG375 &  4200 --  9600     &  9             & LTT7987     \\
\hline
\end{tabular}
\\[1.5ex]
$^a$ JD $-$ 2\,453\,000.0\quad
$^b$ Relative to $B$-band maximum (JD $=2\,453\,482.6 \pm 0.5$).\quad
$^c$ Full-width at half maximum (FWHM) of isolated, unblended night-sky lines.\\[1.8ex]
DUP = Las Campanas 2.5\,m du Pont Telescope + WFCCD; \,http:/$\!$/www.lco.cl/telescopes-information/irenee-du-pont/\\
Caha = Calar Alto 2.2\,m Telescope + CAFOS SiTe; \,http:/$\!$/www.caha.es/CAHA/Instruments/CAFOS/\\
TNG = 3.58\,m Telescopio Nazionale Galileo + DOLORES; \,http:/$\!$/www.tng.iac.es/instruments/lrs/\\
VLT = ESO 8.2\,m Very Large Telescope UT1 + FORS2; \,http:/$\!$/www.eso.org/instruments/fors2/
\end{footnotesize}
\end{table*}

Details of the spectroscopic observations of SN~2005bl are reported in 
Table~\ref{spectra}. 
All two-dimensional spectroscopic frames were first debiased and flat-fielded, 
before an optimal, variance-weighted extraction of the spectra \citep*{Horne86,
IRAFspec} was performed using the {\sc iraf} routine {\sc apall}. Wavelength 
calibration was accomplished with the help of arc-lamp exposures or, whenever 
this was not possible, using the night-sky lines. The instrumental response 
functions required for flux calibration were determined from observations 
of the spectrophotometric standard stars reported in Table~\ref{spectra}. 
Whenever no standard had been observed, the sensitivity curve obtained on 
a different night with the same instrumental configuration was used. 
Atmospheric extinction correction was applied using tabulated extinction 
coefficients for each telescope site. Telluric features were identified in 
the spectra of the spectrophotometric standard stars and removed from the SN 
spectra. To check the calibration, the spectroscopic fluxes were transformed 
into magnitudes by integrating the spectra convolved with \citet{Bessell90} 
filter functions. Whenever necessary, the spectral fluxes were adjusted to 
match the contemporaneous photometry. Finally, spectra of similar quality 
obtained during the same night were combined to increase the signal-to-noise 
ratio (S/N); if the wavelength range of these spectra was different, they 
were averaged in their overlap region.

As a consequence of charge-transfer-efficiency problems of the DOLORES CCD, it 
was not possible to remove the night-sky emission in the TNG spectra cleanly. 
A pattern of negative and positive residuals was left, sometimes at wavelengths 
coinciding with spectral features of the SN, thus limiting the reliability of 
line-depth measurements. To mitigate this problem, the lines affected were 
fitted with polynomials, excluding from the fit the regions of strongest 
residuals, and then the depth was determined in the polynomial curves.

\section{Distance, extinction and host galaxy properties}
\label{Distance and extinction}

Like most 91bg-like SNe~Ia \citep{Gallagher05}, SN~2005bl exploded in an 
early-type host, the elliptical galaxy NGC~4070. Other prominent examples are 
SN~1991bg itself and SN~1997cn in elliptical hosts \citep{Filippenko92b,
Leibundgut93,Turatto96,Turatto98}, and SN~1998de whose host galaxy was of type 
S0 \citep{Modjaz01}.\footnote{SN~1999by \citep{Garnavich04} was hosted by 
NGC~2841, a spiral galaxy, but with the spectral appearance of an elliptical 
galaxy \citep{Gallagher05}.} Since early-type galaxies are assumed to have 
experienced no significant star formation over long times, this behaviour might 
be indicative of 91bg-like SNe~Ia originating from an old stellar population. 
Even if one allows for some more recent star formation in elliptical galaxies 
(e.g. triggered by mergers or interaction), the relative paucity of 91bg-like 
SNe in late-type, actively star-forming galaxies clearly disfavour young or 
intermediate-aged stellar progenitors or progenitor systems.

\begin{figure}   
   \centering
   \includegraphics[width=4.2cm]{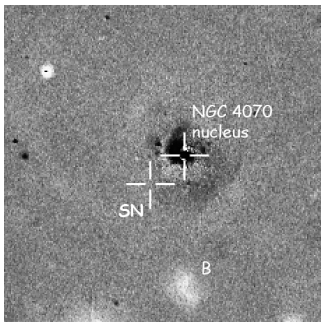}
   \caption{$B-R$ difference image of NGC~4070, constructed as described 
   in the text. Dark shades in the figure correspond to red areas. The field 
   of view is 2 $\times$ 2 arcmin$^2$; north is up, east to the left. 
   Crosshairs mark the explosion site of SN~2005bl and the centre of NGC~4070. 
   `B' denotes a bluish patch, probably a companion galaxy of NGC~4070 (see 
   discussion in the text).}
   \label{fig:BminR}
\end{figure}

A widely accepted paradigm for elliptical galaxies is that they have only 
little interstellar gas and dust (but see \citealt{Patil07} for a more 
sophisticated picture). This is consistent with the negligible host-galaxy 
extinction found in SNe~1991bg, 1997cn and 1998de. However, recent mergers or 
interaction may alter this picture, as exemplified by the radio galaxy Centaurus 
A, the host of SN~1986G \citep{Phillips87}. In this case, the SN lay behind a 
prominent dust lane, and consequently was strongly extinguished. In NGC~4070 no 
such dust lane is visible even in deep images, but SN~2005bl showed signs of 
extinction within its host galaxy, the most prominent being a narrow interstellar 
Na\,{\sc i}\,D line in the spectra at the redshift of the host, with an 
equivalent width (EW) of $2.6 \pm 0.3$\,\AA. 

Since elliptical galaxies lack H\,{\sc ii} regions with strong emission 
lines, their surface colours are good tracers of the internal dust distribution 
\citep{Patil07}. Hence, to investigate the dust content in NGC~4070, we 
constructed a $B-R$ image of the galaxy from the templates obtained with CAFOS 
on UT 2006 March 28, following largely the prescription of \citet{Patil07}: 
after the usual pre-reduction, the $B$- and $R$-band images were spatially 
aligned, and the sky background was subtracted. The images were then scaled to 
contain the same flux inside an aperture of $45$ arcsec around the centre of 
NGC~4070, and subtracted one from the other. 
The difference image is shown in Fig.~\ref{fig:BminR}. Both the centre of the 
galaxy and the position of SN~2005bl are marked. Dark shades correspond to 
redder areas, indicative of either dust or an intrinsically redder stellar 
population. A red region is present to the immediate east and north-east of 
the nucleus, and another more extended but less opaque arc to its south-west. 
This asymmetric surface-colour distribution suggests that there probably is 
dust in NGC~4070. However, at the exact position of SN~2005bl no major blue 
or red structures can be discerned, so that the dust obscuring the SN is 
probably too locally confined to be resolved.

\begin{figure}   
   \centering
   \includegraphics[width=8.3cm]{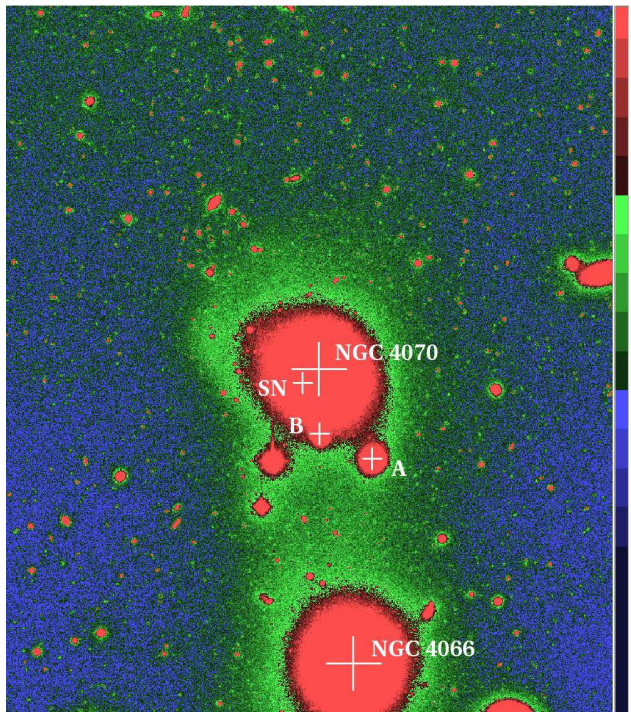}
   \caption{Deep $7.7 \times 9.0$ arcmin$^2$ exposure of NGC~4070 and its 
   neighbour NGC~4066 obtained on UT 2006 March 28 with CAFOS. North is up, 
   east to the left; brightness increases from blue over green to red. The 
   distorted shape of NGC~4070 and the bridge of luminous matter connecting 
   the two galaxies are discernible. `A' and `B' mark companion galaxies of 
   NGC~4070.}
   \label{fig:long_exp}
\end{figure}

Deep images of NGC~4070 reveal some deviation from a perfectly spherical or 
ellipsoidal shape (Fig.~\ref{fig:long_exp}). This is an indication of fairly 
recent interaction, either with the galaxy 2MASX J12040831+2023280 about $1.3$ 
arcmin to its south-west (labelled `A' in Fig.~\ref{fig:long_exp}), or with a 
small knot about $0.8$ arcmin to its south (labelled `B', and most easily seen 
as irregularly-shaped bright patch in the lower part of Fig.~\ref{fig:BminR}). 
Furthermore, in Fig.~\ref{fig:long_exp} a faint, broad bridge of luminous matter 
between NGC 4070 and its equally massive elliptical neighbour galaxy NGC~4066 
can be detected.\footnote{The situation is somewhat reminiscent of SN~Ia 2005cf 
\citep{Pastorello07a}, with the difference that this SN was directly located in 
the tidal bridge.} The two galaxies are offset by about $3.74$ arcmin, which 
corresponds to a projected distance of $114$~kpc (for $H_0=72\,\rmn{km}\,
\rmn{s}^{-1}\rmn{Mpc}^{-1}$).

The exact amount of dust extinction towards SN~2005bl is quite difficult to 
determine, and it constitutes the main uncertainty in the calibration of the 
SN absolute magnitudes. The Galactic component is small, with a colour excess 
$E(B\!-\!V) \approx 0.03$~mag \citep{Schlegel98}. However, as mentioned 
before, the contribution of dust in the host galaxy is significant. 
With EW(Na\,{\sc i}\,D)$= 2.6 \pm 0.3$\,\AA\ and applying 
\begin{displaymath}
E(B\!-\!V)\, =\, 0.16\times{\rm EW}\textrm{(Na\,{\sc i}\,D)}\,,
\end{displaymath}
\citep*{Turatto_proc}, we obtain $E(B\!-\!V)_\mathrm{host} = 0.42 \pm 0.05$~mag. 
However, the lack of knowledge of the extinguishing material's gas-to-dust 
ratio makes this method very uncertain. Therefore, in the case of Type Ia SNe, 
alternative ways to determine the colour excess using the light and colour curves 
are usually preferred. Among the most well-known is the \citet{Lira95} relation, 
which assumes a uniform $B-V$ colour evolution of SNe~Ia between $30$ and $90$\,d 
after maximum (but see \citealt{Wang07} for a recent warning about the use of 
this method). Apparently, this relation seems to hold also for underluminous 
SNe~Ia, which are characterised by much redder $B-V$ colours at maximum and soon 
thereafter. However, the decreasing quality of the SN~2005bl photometry after 
$+30$\,d and the lack of $B$-band observations later than $+42$\,d, result in 
a relatively large range of colour excesses consistent with the Lira relation, 
from $E(B\!-\!V)_\mathrm{host}\approx0.12$ to $0.25$~mag.

Since SN~2005bl is an underluminous SN~Ia with $\Delta$m$_{15}(B)_\mathrm{true}
=1.93$ (cf. Section~\ref{Filter light curves}), SNe~1991bg and 1999by 
($\Delta$m$_{15}(B)_\mathrm{true}=1.94$ and $1.90$, respectively; cf. 
Section~\ref{Discussion}) are natural comparison objects. Despite their numerous 
similarities, the reddening-corrected colour curves of the two latter objects 
differ especially at early phases, SN~1991bg being $\sim$\,$0.15$ mag redder in 
$B-V$ than SN~1999by. This intrinsic colour difference directly propagates to an 
uncertainty in the inferred colour excess of SN~2005bl determined on the basis 
of such comparison. Matching the colours of SN~2005bl to SN~1991bg, we obtain 
$E(B\!-\!V)_\mathrm{host} \approx 0.13 \pm 0.05$ mag, whereas a comparison with 
SN~1999by yields $E(B\!-\!V)_\mathrm{host} \approx 0.29 \pm 0.05$~mag. 

A study of the extinction law towards SN~2005bl, analogous to that presented 
by \citet{Elias-Rosa06} and based on extracting an extinction curve from a 
comparison with coeval spectra of SNe~1991bg and 1999by, yields a host-galaxy 
colour excess of $0.22 \pm 0.02$~mag. Furthermore, no deviation of the 
total-to-selective-extinction parameter $R_V$ from the canonical value of 
$3.1$ as inferred from ``standard'' dust in the Milky Way can be discerned. 

For the rest of the discussion we assume $E(B\!-\!V)_\mathrm{host}=0.17 \pm 
0.08$ mag. This estimate is based solely on the study of the SN colours (giving 
the strongest weight to the comparison with SN~1991bg), and ignores the larger 
colour-excess estimate from the interstellar Na\,{\sc i}\,D line, since the 
latter would result in too blue a colour and too high an absolute luminosity 
given SN~2005bl's spectrophotometric similarity to SNe~1991bg and 1999by. This 
choice of $E(B\!-\!V)_\mathrm{host}$, together with the foreground reddening of 
$0.03$~mag, yields a total colour excess $E(B\!-\!V)_\mathrm{total}= 0.20 \pm 0.08$ 
mag and, adopting a standard \citet*{Cardelli89} reddening law with $R_V = 3.1$, 
a total $B$-band extinction along the line of sight of $A_B = 0.82 \pm 0.33$~mag.

With a recession velocity corrected to the CMB reference frame of $7534 \pm 33$ 
km\,s$^{-1}$ (Table~\ref{properties}), NGC~4070 is well within the Hubble flow. 
Adopting $H_0=72\,\rmn{km}\,\rmn{s}^{-1}\rmn{Mpc}^{-1}$ \citep{Freedman01,
Spergel03}, this corresponds to a distance of $104.6$ Mpc and a kinematical 
distance modulus $\mu = 35.10$~mag, similar to that of the Coma Cluster. 
However, the latter is about 14\degr45\arcmin\ away (which, at the given 
distance, corresponds to $27$ Mpc), excluding any physical association.  
Accounting for an uncertainty of $300$ km\,s$^{-1}$ arising from a possible 
peculiar motion of NGC~4070, we obtain $\mu = 35.10 \pm 0.09$~mag.

\section{Photometric evolution}
\label{Photometric evolution}

\subsection{Filtered light curves}
\label{Filter light curves}

\begin{figure}   
   \centering
   \includegraphics[width=8.4cm]{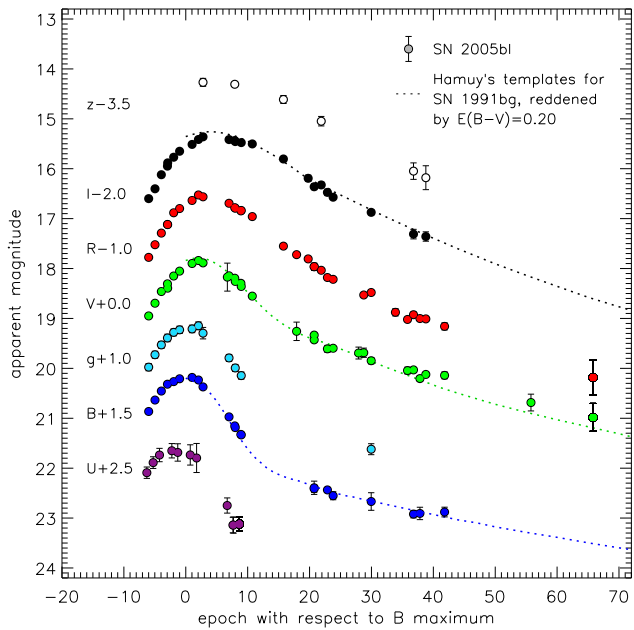}
   \caption{$U\!BV\!RI$ Bessell and $gz$ Sloan light curves of SN~2005bl. The 
   $BV\!RI$ data (Table~\ref{SN_mags1}) are $S$- and $K$-corrected except for 
   the latest phases, while the $U$, $g$ and $z$ data (Tables~\ref{SN_mags1}, 
   \ref{SN_mags2} and Contreras et~al. in prep.) are not. Hamuy et al.'s 
   (\citeyear{Hamuy96c}) templates for SN~1991bg are shown for comparison 
   (dotted lines).}
   \label{fig:UBVRIgz}
\end{figure}

The photometric observations of SN~2005bl are among the earliest ever obtained 
for a 91bg-like SN. In Fig.~\ref{fig:UBVRIgz} we present the Bessell $U\!BV\!RI$ 
and Sloan $gz$ light curves, i.e. the data of Tables~\ref{SN_mags1} and 
\ref{SN_mags2}, and Contreras et~al. (in prep.). $S+K$-corrections have been 
applied to the $BV\!RI$ bands. Also shown are $BV\!I$-templates constructed 
from SN~1991bg \citep{Hamuy96c}, reddened by $E(B-V)=0.20$ mag. These provide 
an excellent match to SN~2005bl.

Compared to normal-luminosity SNe~Ia the light curves of SN~2005bl are 
characterised by a fast rise to and decline from the light-curve peak, especially 
in the blue bands. Moreover, in the $B$ band the settling to the exponential tail 
(corresponding to the bend parameter $t_\mathrm{b}$ of \citealt{Pskovskii84} and 
the intersection parameter $t_2^B$ as defined by \citealt{Hamuy96c}) occurs at 
a remarkably early epoch, only about $15$\,d after maximum, as compared to 
$25$--$38$\,d for intermediate and slow decliners \citep{Hamuy96c}. Consequently, 
the decline from the peak to the onset of the radioactive tail is only 
$\sim$\,$1.9$ mag, less than in most SNe~Ia with shallower initial decline. 

The exact value of the \citet{Phillips93} decline rate parameter 
$\Delta$m$_{15}(B)$ is difficult to measure directly because of a gap in 
the light curves around $+15$\,d. A polynomial fit including the $B$-band data 
up to one month past maximum yields $\Delta$m$_{15}(B)=1.91$, and we adopt a 
conservative error of $0.10$ to account for the fact that the fit around $+15$\,d 
is not very well constrained. After correction for the mitigating effects of 
extinction \citep{Phillips99} this turns into an actual decline rate 
$\Delta$m$_{15}(B)_\mathrm{true} = 1.93 \pm 0.10$.

Besides the $B$ band, differences with respect to normal-luminosity SNe~Ia are 
most pronounced in the near-IR ($I$ and $z$ bands), where -- like in other 
91bg-like SNe -- the secondary light-curve maximum is absent, and the main 
maximum is delayed with respect to that in $B$ (JD $2\,453\,482.6 \pm 0.5$) 
by a couple of days rather than advanced. In general, the instance of peak 
brightness seems to be the more delayed the redder the band is. $V$-band maximum 
occurs $2.3$\,d after that in $B$, and those in $R$ and $I$ are delayed by $3.3$ 
and $4.4$\,d, respectively (see Table~\ref{properties}), whereas in $U$ the 
light-curve peak precedes that in $B$ by about $1.0$\,d. For the $z$ band the 
sparse photometric coverage does not allow for an exact determination of the 
time of maximum light, but it can be estimated to be similarly delayed as in 
$R$ or $I$.

The absolute peak magnitudes of SN~2005bl (Table~\ref{properties}), calculated 
adopting the distance and extinction estimates presented in Section~\ref{Distance 
and extinction}, reveal that the SN is underluminous by $1$ to $2$ mag in all 
filters compared to a canonical, $\Delta$m$_{15}(B)_\mathrm{true} = 1.1$ SN~Ia, 
the difference being most pronounced in the blue bands.

\subsection{Bolometric light curve}
\label{Bolometric light curve}

\begin{figure}   
   \centering
   \includegraphics[width=8.4cm]{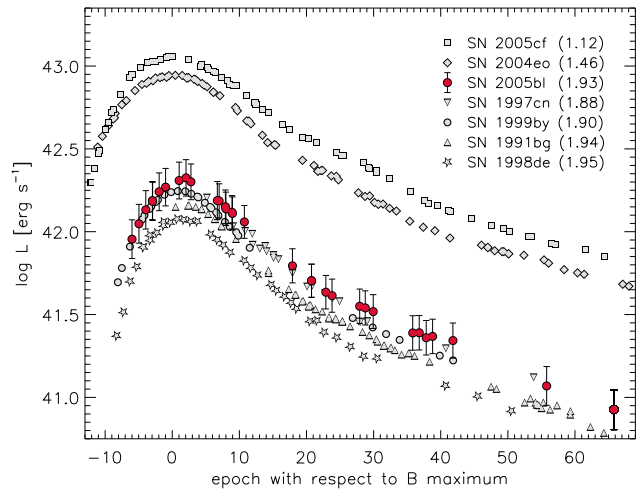}
   \caption{Quasi-bolometric light curves of SNe~2005bl, 1991bg, 1997cn, 1998de, 
   1999by, 2004eo and 2005cf, obtained by integrating the $U$-through-$I$-band 
   fluxes (for the adopted distance and extinction parameters see 
   Table~\ref{SNe_faint} and \citealt{Pastorello07a,Pastorello07b}). 
   Error bars are shown for SN~2005bl only, and account for uncertainties in 
   the photometric calibration, distance and extinction estimates. The 
   $\Delta$m$_{15}(B)_\mathrm{true}$ of the SNe is given in parentheses.}
   \label{fig:bolom}
\end{figure}

The differences in luminosity can clearly be seen in Fig.~\ref{fig:bolom}, where 
the quasi-bolometric light curve of SN~2005bl is compared to those of the other 
underluminous SNe~Ia 1991bg, 1997cn, 1998de and 1999by \citep{Filippenko92b,
Leibundgut93,Turatto96,Turatto98,Modjaz01,Garnavich04}, and the intermediate 
decliner SN~Ia 2004eo ($\Delta$m$_{15}(B)_\mathrm{true}=1.46$, 
\citealt{Pastorello07b}), which forms a bridge to canonical, normal-luminosity 
SNe~Ia such as 2005cf ($\Delta$m$_{15}(B)_\mathrm{true}=1.12$, 
\citealt{Pastorello07a}). The integrated optical light curves 
\citep*[see, e.g.,][]{Nomoto90} were constructed in the following way: in a 
first step the $U$-through-$I$ magnitudes were converted to monochromatic fluxes 
and the spectral energy distribution (SED) was interpolated linearly. The SED was 
then integrated over frequency, assuming zero flux at the integration limits, 
which are given by the blue edge of the $U$ band and the red edge of the $I$ 
band. Whenever no $U$-band observations were available, or the coverage of this 
band was incomplete, a correction derived from SN~1999by was applied to the 
analogously constructed $B$-through-$I$ light curve. This method appeared more 
reliable than applying $U$-band corrections based on ordinary SNe~Ia. 

As Fig.~\ref{fig:bolom} shows, 91bg-like SNe form a fairly homogeneous group 
in terms of bolometric light-curve shape and luminosity. Both their light-curve 
width and their luminosities distinguish them even from intermediate decliners 
such as SN~2004eo. SNe~Ia with $\Delta$m$_{15}(B)_\mathrm{true}$ $\sim$\,$1.50$--$1.85$ 
(which are rare, cf. Section~\ref{Photometric behaviour of underluminous SNe Ia}) 
would probably fall in the gap between SN~2004eo and the 91bg-like SNe. In 
Fig.~\ref{fig:bolom} they are not included because their light-curve coverage 
is insufficient in some of the relevant bands. Within the 91bg-like group, 
SN~2005bl appears to be the brightest object ($\log L_\mathrm{max} = 42.31 \pm 
0.11$), but the differences are mostly within the error bars, which for 
SN~2005bl are dominated by the uncertainty in the host-galaxy extinction.

In the only rapidly-declining SN~Ia with extended near-IR photometry, SN~1999by 
\citep{Hoeflich02,Garnavich04}\footnote{SN~1991bg has been observed in $JHK$ 
by \citet{Porter92} (five epochs), but only the three epochs close to maximum 
light have been calibrated and published \citep{Krisciunas04}}, the $JHK$ bands 
contribute $\sim$\,25\,$\%$ to the total bolometric flux around maximum, and 
$\sim$\,36\,$\%$ after one month. The corresponding numbers measured for 
SN~2004eo are $\sim$\,16\,$\%$ and $\sim$\,37\,$\%$, respectively. This means 
that at early phases the near IR gives a larger contribution in the fast 
decliners than it does in intermediate and slow decliners, but that by $+30$\,d 
the difference has vanished. 
Wavelength regions other than the optical and near-IR appear to contribute very 
little to the total bolometric flux in SNe~Ia \citep*{Suntzeff96,Contardo00}. 
Therefore, neglecting these regimes should not cause a significant underestimate 
of the true bolometric luminosity.

\subsection{Colour evolution}
\label{Colour evolution}

\begin{figure}   
   \centering
   \includegraphics[width=8.4cm]{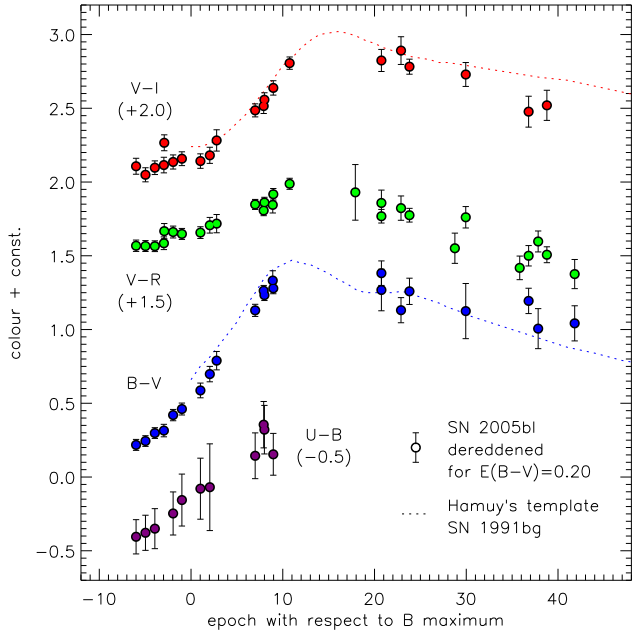}
   \caption{Time-evolution of the $U-B$, $B-V$, $V-R$ and $V-I$ colour indices of 
   SN~2005bl. The curves have been reddening-corrected adopting a \citet{Cardelli89} 
   extinction law with $E(B-V)=0.20$ mag and $R_V=3.1$.}
   \label{fig:colours}
\end{figure}

Fig.~\ref{fig:colours} presents the time-evolution of the $U-B$, $B-V$, $V-R$ 
and $V-I$ colours of SN~2005bl. The good agreement with the colour curves of 
SN~1991bg \citep{Hamuy96c} also shown in the figure is evident. Except for the 
late $V-R$ points, all colour indices are $>0$ throughout the investigated 
period, indicating that SN~2005bl -- like other rapidly-declining SNe~Ia -- 
was a rather red event. In particular it was much redder at early phases than 
ordinary SNe~Ia, which are characterised by a $B-V$ between $0.0$ and $-0.1$ 
at maximum light. Remarkably, the evolution of SN~2005bl's various colour 
indices with time is very similar. From our first observations at $-6$\,d on, 
all colours become monotonically redder until at least $\sim$\,$10$\,d after 
maximum, and almost simultaneously exhibit a red peak between $+12$ and 
$+17$\,d, followed by a monotonic bluening which probably lasts beyond the end 
of our photometric coverage around $+40$\,d. Such a high degree of similarity 
between different colour indices is not encountered in normal-luminosity 
SNe~Ia (cf. Section~\ref{Photometric behaviour of underluminous SNe Ia}).

\section{Spectroscopic evolution}
\label{Spectroscopic evolution}

\subsection{Spectra of SN~2005bl}
\label{Spectra of SN 2005bl}

\begin{figure}   
   \centering
   \includegraphics[width=8.4cm]{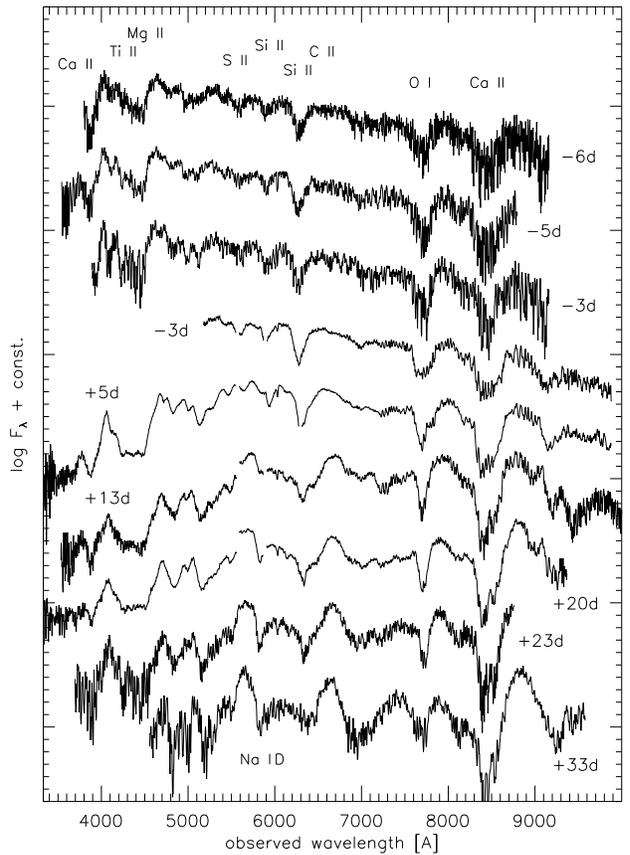}
   \caption{Time sequence of SN~2005bl spectra in the observer frame. 
   The phases reported next to each spectrum are with respect to $B$-band 
   maximum. The $-3$\,d (blue part) and $+33$\,d spectra have been smoothed 
   using kernel sizes of $600$\,km\,s$^{-1}$. Gaps in the TNG spectra 
   indicate the positions of the strongest night-sky residuals, which have 
   been cut in order to avoid confusion with true SN or host-galaxy features 
   (see Section~\ref{Spectroscopic data} and Table~\ref{spectra}).}
   \label{fig:2005bl_spectra}
\end{figure}

The spectra of SN~2005bl presented in Fig.~\ref{fig:2005bl_spectra} cover 
the interval from $6$\,d before to $33$\,d after $B$-band maximum light, i.e., 
the photospheric phase and the early transition stages towards the nebular 
phase. During this entire period, the SN follows the evolution of other SNe~Ia, 
in particular of the 91bg-like subclass. The pre-maximum spectra have a blue 
continuum, with characteristic P-Cygni lines of Si\,{\sc ii}, S\,{\sc ii}, 
Ca\,{\sc ii} and Mg\,{\sc ii} superimposed. With respect to normal-luminosity 
SNe~Ia, additional strong Ti\,{\sc ii} absorptions are visible, and the 
Si\,{\sc ii} $\lambda5972$ and O\,{\sc i} $\lambda7774$ absorptions are more 
pronounced. Shortly after maximum light the blue flux decreases significantly, 
and also the  S\,{\sc ii} and Si\,{\sc ii} $\lambda5972$ lines fade rapidly, 
and are no longer detectable two weeks after maximum. Si\,{\sc ii} $\lambda6355$ 
is, like most other photospheric lines, somewhat more persistent, and can still 
be discerned beyond $+20$\,d. Na\,{\sc i} D shows a trend opposite to that of 
Si and S, first being just visible as a shoulder in the blue wing of the 
Si\,{\sc ii} $\lambda5972$ line on day $+5$, but evolving to a distinct 
absorption feature by day $+20$. In the same period, Fe\,{\sc ii} emission 
lines start to dominate the spectrum.

\subsection{Spectroscopic comparison with other underluminous SNe~Ia}
\label{Spectroscopic comparison}

\begin{figure}   
   \centering
   \includegraphics[width=8.4cm]{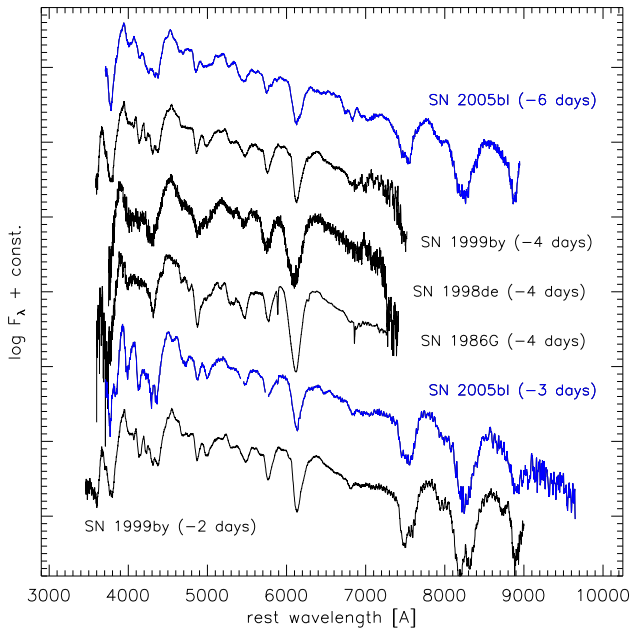}
   \caption{Comparison of pre-maximum spectra of underluminous SNe~Ia. The 
   spectra were reddening-corrected according to the $B-V$ colour excesses 
   reported in Section~\ref{Discussion}. The $-6$\,d spectrum and the blue 
   part of the combined $-3$\,d spectrum of SN~2005bl were boxcar smoothed 
   using kernel sizes of 2300 and 3400\,km\,s$^{-1}$, respectively.} 
   \label{fig:comp_minus05}
\end{figure}

\begin{figure}   
   \centering
   \includegraphics[width=8.4cm]{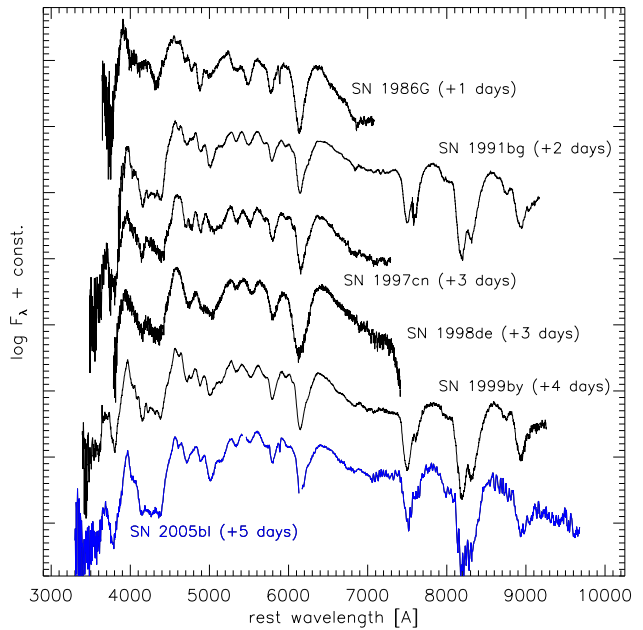}
   \caption{The same as Fig.~\ref{fig:comp_minus05}, but some days after 
   maximum light in $B$.} 
   \label{fig:comp_plus04}
\end{figure}

\begin{figure}   
   \centering
   \includegraphics[width=8.4cm]{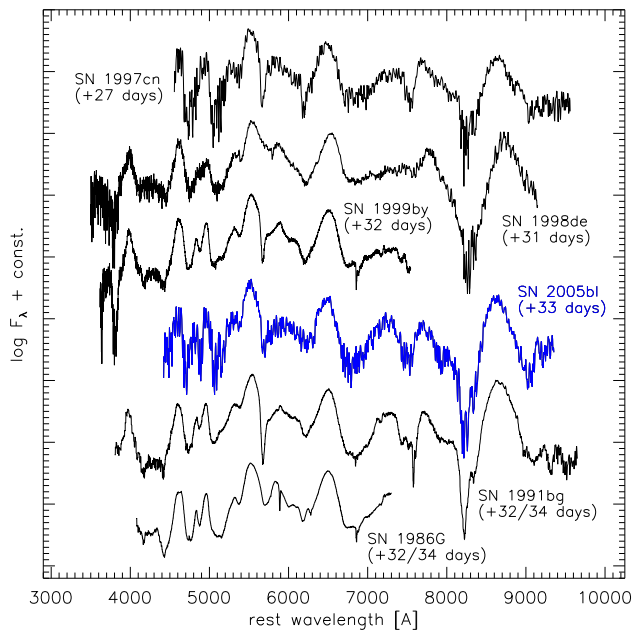}
   \caption{The same as Fig.~\ref{fig:comp_minus05}, but 4--5 weeks after 
   maximum light in $B$. The spectra of SNe~1997cn and 2005bl were smoothed 
   using kernel sizes of $600$\,km\,s$^{-1}$.}
   \label{fig:comp_plus33}
\end{figure}

\begin{figure}   
   \centering
   \includegraphics[width=8.4cm]{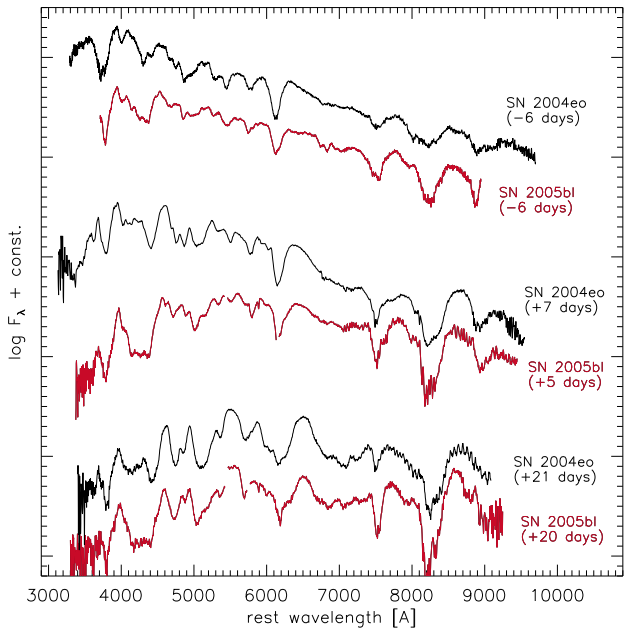}
   \caption{Spectroscopic comparison of SN~2005bl and the intermediate decliner 
   SN~2004eo \citep{Pastorello07b} at three different epochs. The spectra of 
   SN~2004eo have been dereddened for $E(B-V)=0.109$ mag; the $-6$\,d spectrum 
   of SN~2005bl was smoothed using a kernel of 2300\,km\,s$^{-1}$.}
   \label{fig:comp_2004eo}
\end{figure}

In Figs.~\ref{fig:comp_minus05} to \ref{fig:comp_plus33} the spectra of 
SN~2005bl are compared with those of other underluminous SNe~Ia at epochs 
of about $-5$\,d, $+4$\,d and $+30$\,d, respectively. For illustration, 
a comparison with the spectra of the intermediate decliner SN~Ia 2004eo 
($\Delta$m$_{15}(B)_\mathrm{true}=1.46$, \citealt{Pastorello07b}) is made 
in Fig.~\ref{fig:comp_2004eo}. The figures show that 91bg-like SNe form a 
relatively homogeneous spectroscopic subclass, distinct from SNe~Ia with 
normal luminosity and even from those with $\Delta$m$_{15}(B)_\mathrm{true}$ 
close to $1.5$. 

Already at $-5$\,d (Fig.~\ref{fig:comp_minus05}) the underluminous SNe~Ia 
show evident Ti features between $4000$ and $4400$\,\AA, which are absent in 
SN~2004eo (Fig.~\ref{fig:comp_2004eo}) where this region is dominated by Si 
and Mg lines. At the same time Fig.~\ref{fig:comp_2004eo} suggests that at 
$-6$\,d the continuum temperatures are not too different, indicating that the 
strength of the Ti features in 91bg-like SNe may not be a pure temperature 
effect but require a truly larger Ti abundance (see also Section~\ref{Parameters 
and composition}). In SN~1986G \citep{Phillips87} the Ti lines are less 
pronounced than in SNe~1998de \citep{Matheson07}, 1999by \citep{Garnavich04} 
and 2005bl, emphasising its transitional character between ``normal'' and 
strictly 91bg-like SNe~Ia. Also, the absolute depth of the Si\,{\sc ii} 
$\lambda6355$ line in SN~1986G resembles more that of SN~2004eo than those of 
SNe~1999by and 2005bl. The Si\,{\sc ii} lines and most other features of 
SN~1998de at day $-5$ are broader and at bluer wavelength than in the other 
SNe of the sample. The spectra of SNe~2005bl and 1999by are essentially 
identical, the only exception being the flux depletion in SN~2005bl redwards 
of the Si\,{\sc ii} $\lambda6355$ line at $-6$ and $-5$\,d, which we attribute 
to C\,{\sc ii} $\lambda6580$ (see also Section~\ref{Element abundances from 
synthetic spectra}). O\,{\sc i} $\lambda7774$ is particularly pronounced in 
all underluminous SNe~Ia for which the spectral region is covered.

By a few days after maximum light (Fig.~\ref{fig:comp_plus04}) the spectra have 
evolved significantly with the continuum being much redder now, but the degree of 
homogeneity is still remarkably high. The Ti troughs in the blue are now fully 
developed, showing the characteristic flat bottom that distinguishes 91bg-like 
from other SNe~Ia (Fig.~\ref{fig:comp_2004eo}). The W-shaped S\,{\sc ii} lines 
around $5500$\,\AA\ are comparatively weak. Again SN~1986G takes an intermediate 
position, with the characteristic properties of the underluminous class being 
less pronounced than in the other objects shown in Fig.~\ref{fig:comp_plus04}.

The spectra taken 4--5 weeks past maximum (Fig.~\ref{fig:comp_plus33}) reveal 
that at those phases the transition to the nebular phase has already started. 
Emission lines of Fe-group elements are visible, but the pseudo-continuum has 
not yet vanished. Ca\,{\sc ii}, O\,{\sc i} and the Ti trough are still prominent 
in absorption, while S\,{\sc ii} lines cannot be identified any longer. A 
characteristic feature of many 91bg-like SNe~Ia at those epochs is the remarkably 
narrow Na\,{\sc i}\,D absortion near $5700$\,\AA. Especially in SNe~1991bg 
\citep{Filippenko92b,Leibundgut93,Turatto96}, 1997cn \citep{Turatto98} and 
1999by this line is very deep and distinct, while it is less pronounced in 
SNe~2005bl and, in particular, 1986G and 1998de \citep{Modjaz01}, where also 
the narrow core is absent. As already noted by \citet{Modjaz01}, SN~1998de 
deviates significantly from other 91bg-like SNe between $6800$ and $7700$\,\AA, 
exhibiting a relatively smooth continuum without strong O\,{\sc i} $\lambda7774$ 
feature. 

Unfortunately, no late-time spectra of SN~2005bl were obtained, so that it 
cannot be verified whether the distinct, broad nebular emission feature near 
$7300$\,\AA, which characterises SN~1991bg and distinguishes it from ordinary 
SNe~Ia at late epochs \citep{Filippenko92b,Turatto96,Mazzali97}, is also present 
in SN~2005bl.

\subsection{Ejecta velocities}
\label{Ejecta velocities}

Although ejecta velocities inferred from the blueshift of the absorption 
minima of P-Cygni lines suffer from a number of uncertainties (such as the 
typically strong line blending in SN~Ia spectra), they do provide important 
information about the kinetic energy of the ejecta, and can be used as an 
observationally accessible parameter for comparison studies between different 
SNe. Furthermore, the range of velocities encompassed by different elements 
provides insight into the chemical stratification of the ejecta, and hence into 
nucleosynthesis conditions. This fact has recently been made use of in the 
Zorro diagnostics \citep{Mazzali07}. In SNe~Ia, especially the Si\,{\sc ii} 
$\lambda6355$ has proved to be a suitable velocity indicator, since the line is 
usually well pronounced, fairly unblended and well visible over a relatively 
long period. 

\begin{figure}   
   \centering
   \includegraphics[width=8.4cm]{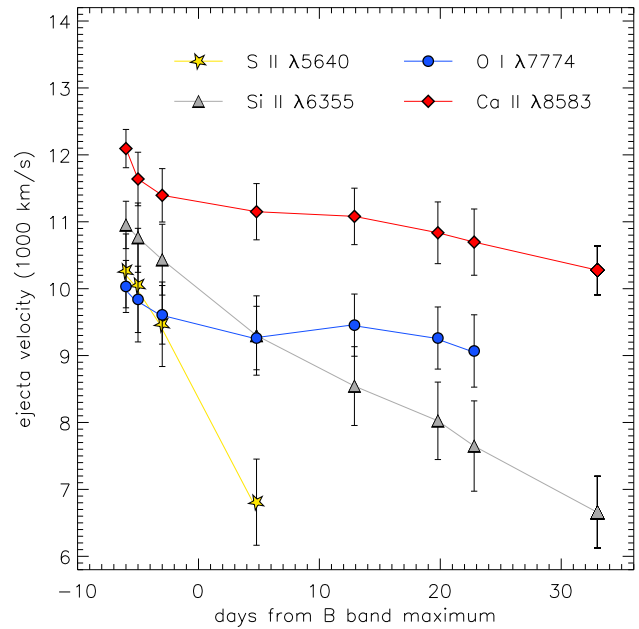}
   \caption{Expansion velocities of S\,{\sc ii} $\lambda5640$, Si\,{\sc ii} $\lambda6355$, 
   O\,{\sc i} $\lambda7774$ and the Ca\,{\sc ii} NIR-triplet as measured from the minima 
   of the P-Cygni line profiles in SN~2005bl.}
   \label{fig:ejecta_velocities}
\end{figure}

The Si line velocity in SN~2005bl (Fig.~\ref{fig:ejecta_velocities}) evolves 
from $\sim\!11\,000$ km\,s$^{-1}$ at $-6$\,d to $\sim\!6600$ km\,s$^{-1}$ at 
$+33$\,d. At maximum light, it is about $10\,000$ km\,s$^{-1}$, which is low 
for a SN~Ia. The velocity evolution of S\,{\sc ii} $\lambda5640$ resembles that 
of Si\,{\sc ii}, but the temporal decrease is steeper, and the absolute values 
are systematically lower by $500$--$2500$ km\,s$^{-1}$, meaning that the line 
predominantly forms in deeper layers of the ejecta (note that the line is 
visible only until $+5$\,d). The Ca\,{\sc ii} IR triplet, on the other hand, 
has higher velocities than Si\,{\sc ii}, and the difference increases from 
about $1000$ km\,s$^{-1}$ at $-6$\,d to $3500$ km\,s$^{-1}$ at $+33$\,d. This 
relatively shallow decrease in velocity means that Ca\,{\sc ii} lines mainly 
form well above the photosphere. Finally, O\,{\sc i} $\lambda7774$ also 
exhibits a fairly constant velocity of $9000$--$10\,000$ km\,s$^{-1}$ during 
the entire period of our observations, but this result is not very robust as 
the feature may be substantially blended with Mg\,{\sc ii} $\lambda7890$.

\section{Spectral modelling}
\label{Spectral modelling}

A 1D Monte-Carlo spectrum synthesis code \citep{Mazzali00} was used to simulate 
the radiation-transport processes in the expanding ejecta of SN~2005bl. Basic 
assumptions include spherical symmetry, a Chandrasekhar-mass explosion, and an 
underlying density profile adopted from the W7 explosion model of 
\citet*{Nomoto84}. No attempt was made to obtain better results altering these 
ingredients. Synthetic fits to four early-time spectra of SN~2005bl were 
obtained and are presented here, after a short introduction to the concept of 
the code and the underlying model.

\subsection{Concept of the radiative transfer code}
\label{Code concept}

Only a brief outline is given here; for detailed descriptions of the code see 
\citet{AbbottLucy85}, \citet{MazzaliLucy93}, \citet{Lucy99} and \citet{Mazzali00}, 
where the basic developments are documented.

The radiative transfer is performed above the photosphere, which is located at 
an adjustable radius $r_\mathrm{ph}$. Energy deposition from radioactivity is 
assumed to occur below the photosphere, from which a blackbody continuum is 
thought to be emitted. In the ``atmosphere'' above, radiative equilibrium is 
supposed to hold. Photons interact with lines and scatter on free electrons, 
but no continuum formation is assumed (Schuster-Schwarzschild approximation).
The underlying W7 density profile is scaled to match the epoch of each spectrum, 
assuming homologous expansion, i.e. $r=v\,t$ for each particle.\footnote{This 
is almost exact for our purposes, as the expansion is homologous already 
$\sim$\,10\,s after explosion onset (see e.g. \citealt{Roepke05}).} Element 
abundances are assumed to be homogeneous inside the envelope. They can be freely 
adjusted in order to match a given observed spectrum.

In our Monte Carlo calculation, radiation packets are followed from their 
emission at the photosphere through their interaction history, until they 
either escape or are reabsorbed at the photosphere. The code takes into account 
scattering processes (on atoms\,/\,ions and electrons) as well as photon branching 
(absorption in an atomic line and subsequent reemission in another). Line optical 
depths are calculated in the Sobolev approximation, applicable for fast-expanding 
atmospheres.

Ionisation and excitation conditions are calculated from the radiation-field 
statistics. This is done using approximate NLTE formulae (see references above), 
which in principle employ LTE radiative rates, but additionally take into account 
the dilution of the radiation field. Collision processes are neglected. A radiation 
temperature $T_\mathrm{R}$, determining radiative rates, is calculated in each 
zone of the envelope. As in \citet{MazzaliLucy93} it is chosen such that the mean 
frequency of the radiation field inside the zone matches that of a blackbody at 
$T_\mathrm{R}$. The matter state and the radiation field are iterated until 
sufficient convergence is achieved. Within this process, the temperature of the 
inner-boundary blackbody spectrum is adjusted in order to match the bolometric 
SN luminosity.

Finally, the emitted spectrum is recalculated solving the formal integral, 
employing source functions obtained from the packet statistics. This yields 
smooth spectra at relatively low packet numbers.

\begin{figure}   
   \centering
   \includegraphics[width=8.4cm]{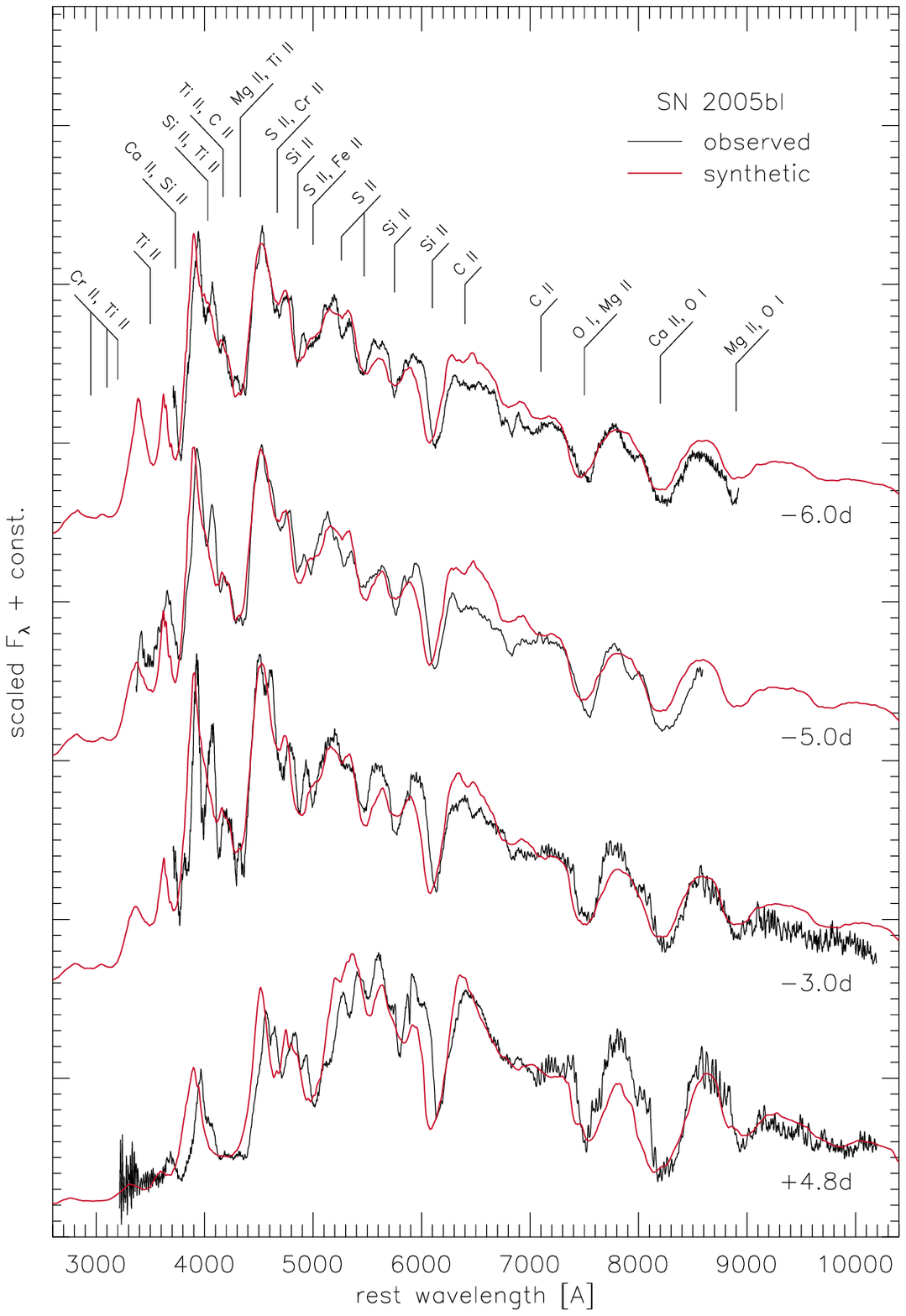}
   \caption{Synthetic fits to the SN~2005bl spectra at days $-6.0$, $-5.0$, 
   $-3.0$ and $+4.8$. The observed spectra were smoothed with kernel sizes of 
   2300 km\,s$^{-1}$ ($-6.0$\,d) and 3400 km\,s$^{-1}$ ($-5.0$\,d and 
   $-3.0$\,d--blue part) for presentation. The values of relevant fit-parameters 
   are summarised in Table~\ref{model parameters}. Atop the $-6$\,d spectrum an 
   identification of the most important lines is given.}
   \label{fig:models}
\end{figure}

\begin{table*}
\caption{Physical parameters of SN~2005bl inferred from synthetic spectra, 
assuming a rise time of $17$\,d in $B$. The mass fractions of selected elements 
are also recorded, and solar photospheric abundances \citep*{Asplund05} are 
shown for comparison.}
\label{model parameters}
\begin{footnotesize}
\begin{tabular}{cccccc}
\hline 
                                       &    $-6.0$\,d      &    $-5.0$\,d      &    $-3.0$\,d      &    $+4.8$\,d      &        solar        \\
\hline
time from explosion                    &    $11.0$\,d      &    $12.0$\,d      &    $14.0$\,d      &    $21.8$\,d      &                     \\ 
bolometric luminosity [erg\,s$^{-1}$]  &$1.24\times10^{42}$&$1.56\times10^{42}$&$2.06\times10^{42}$&$2.56\times10^{42}$&                     \\
photospheric velocity [km\,s$^{-1}$]   &      $7500$       &      $7350$       &      $7100$       &      $6000$       &                     \\
photospheric blackbody temperature [K] &    $10\,620$      &    $10\,790$      &    $10\,670$      &      $9230$       &                \\[1.5ex]
X(C)                                   &     $0.045$       &     $0.030$       &     $0.010$       &     $0.000$       & $2.16\times10^{-3}$ \\
X(O)                                   &     $0.905$       &     $0.878$       &     $0.847$       &     $0.788$       & $5.36\times10^{-3}$ \\
X(Mg)                                  &     $0.015$       &     $0.040$       &     $0.060$       &     $0.080$       & $6.04\times10^{-4}$ \\
X(Si)                                  &     $0.025$       &     $0.037$       &     $0.060$       &     $0.090$       & $6.66\times10^{-4}$ \\
X(S)                                   &     $0.006$       &     $0.009$       &     $0.013$       &     $0.020$       & $3.24\times10^{-4}$ \\
X(Ti)                                  &$3.7\times10^{-4}$ &$7.3\times10^{-4}$ &$1.4\times10^{-3}$ &$4.5\times10^{-3}$ & $2.79\times10^{-6}$ \\
X(Cr)                                  &$3.7\times10^{-4}$ &$7.3\times10^{-4}$ &$1.4\times10^{-3}$ &$4.5\times10^{-3}$ & $1.66\times10^{-5}$ \\
X(stable Fe)                           &$1.0\times10^{-4}$ &$1.2\times10^{-4}$ &$4.0\times10^{-4}$ &$1.4\times10^{-3}$ & $1.15\times10^{-3}$ \\
X($^{56}$Ni + decay products)          &        --         &        --         &$2.0\times10^{-4}$ &$2.5\times10^{-3}$ &                     \\
\hline
\end{tabular}
\end{footnotesize}
\end{table*}

\subsection{Physical parameters and the chemical composition of the models}
\label{Parameters and composition}

The basic parameters and composition inferred from the best-fitting synthetic 
spectra are reported in Table~\ref{model parameters}. In Fig.~\ref{fig:models} 
these models are compared to the observed spectra, and line identifications 
for the $-6$\,d spectrum are shown. A rise time to the $B$-band maximum of 
$17.0$\,d is assumed, slightly less than the fiducial value of $\sim$\,$19.5$\,d 
for normal-luminosity SNe~Ia \citep{Riess99,Conley06}. This choice is discussed 
in Section~\ref{Rise time}.

While the overall shape of the continua is nicely matched and most features 
are well reproduced, the main shortcomings of the models are a flux excess 
in the emission component of Si\,{\sc ii} $\lambda6355$ at early times, and, 
more worrysome, a mismatch in the position of some lines. Additionally, many 
synthetic features are too broad and strongly blended, as they were in models 
for SN~1991bg \citep{Mazzali97}. No substantial improvement is obtained 
reducing the photospheric velocity $v_\mathrm{ph}$, which indicates that the 
W7 density profile may not be perfectly suitable for 91bg-like SNe, and that a 
steeper density gradient may be required. With $v_\mathrm{ph} = 7500$ km\,s$^{-1}$ 
$6$\,d prior to maximum light, the inferred photospheric velocity is 
significantly lower than in normal SNe~Ia ($\sim\!10\,000$ to $11\,000$ 
km\,s$^{-1}$ at comparable epochs). By day $+5$, $v_\mathrm{ph}$ further decreases 
by about $1500$ km\,s$^{-1}$. Singly and doubly ionised species dominate the 
ejecta. Especially the singly ionised species leave strong imprints on the 
spectra, as can be seen from the line identification in Fig.~\ref{fig:models}.
In all spectra modelled here the heavy-element content is comparatively low, 
but Table~\ref{model parameters} shows that it increases with time, indicating 
composition stratification in the ejecta.\footnote{For this reason it would 
be necessary to work with the stratified version of the code \citep{Stehle05} 
if accurate abundances were required. This reaches beyond the scope of this 
paper. Therefore, the composition reported in Table~\ref{model parameters} 
reflects the overall trend, but the absolute numbers should be taken with 
caution.} At the higher velocities the ejecta are almost entirely made up of 
unburned material, whereas NSE elements are essentially absent. In particular, 
no Ni and Co, and only very little Fe are included in the $-6$ and $-5$\,d 
models. After maximum light, IMEs and Fe-group elements become more abundant, 
but still less so than in normal-luminosity SNe~Ia \citep[cf., e.g.,][]{Stehle05}. 
Interestingly, Ti and Cr, whose abundances can be well constrained from the 
depth of the characteristic trough between $4000$ and $4400$\,\AA, appear to 
be more abundant than Ni, Co and Fe, yet another indication for explosion 
conditions which disfavour burning to NSE.

\section{Discussion}
\label{Discussion}

In a number of works \citep[e.g.][]{Filippenko92b,Leibundgut93,RuizLapuente93,
Turatto96,Mazzali97,Turatto98,Modjaz01,Garnavich04} 91bg-like SNe~Ia have been 
studied and found to be different from normally luminous SNe~Ia in several 
respects. Besides their lower luminosity, they have rapidly declining light 
curves which do not show the characteristic secondary maximum in the near-IR, 
cooler spectra with significant Ti\,{\sc ii} absorption troughs, low ejecta 
velocities and large values of the $\mathcal{R}$(Si) parameter \citep{Nugent95}. 
Nevertheless, the light-curve width and peak luminosity of underluminous SNe~Ia 
seem to obey a correlation \citep{Garnavich04}, but not the same one as more 
ordinary SNe~Ia. Also, 91bg-like SNe seem to fit smoothly into the Zorro plot 
\citep{Mazzali07}, where they represent the extremely $^{56}$Ni-poor, IME-rich 
end of the SN Ia distribution (but see Section~\ref{Element abundances from 
synthetic spectra} for a revised picture). With the increasing number of 
well-observed objects of this class, it is interesting to revisit some of the 
aspects mentioned above and to investigate in more detail the degree of 
homogeneity, or rather diversity, that these SNe exhibit. In Table~\ref{SNe_faint} 
elementary information on our comparison sample of rapidly-declining SNe~Ia is 
collected; selection criteria were a reasonably dense light-curve coverage 
starting at least around maximum light, and $\Delta m_{15}(B)_\mathrm{true}>1.5$.

\begin{table*}
\caption{Comparison sample of SNe~Ia with $\Delta m_{15}(B)_\mathrm{true}>1.5$, ordered by increasing $\Delta m_{15}(B)_\mathrm{true}$.}
\label{SNe_faint}
\begin{footnotesize}
\begin{tabular}{lccccclccl}
\hline 
\ \ SN & $\!\!\!\!\Delta m_{15}(B)_\mathrm{true}\!\!\!\!$ & M$_{B,\mathrm{max}}$ & M$_{V,\mathrm{max}}$ & $\mathcal{R}$(Si)$_\mathrm{max}$ & $\dot v^a$ & $E(B\!-\!V)^b$ & $\mu^c$ & morph.$^d$ & references         \\
\hline
1990af &       1.57(0.05)       & $-18.96(0.25)$ & $-18.98(0.19)$ &                             &          & 0.07(0.06)   & 36.59(0.05) & SB0         & H96a, H96b, P99     \\
2000dk &       1.57(0.09)       & $-18.84(0.18)$ & $-18.83(0.17)$ &                             &          & 0.07(0.03)   & 34.18(0.13) & E           & J06, VSNET         \\
1999gh &       1.69(0.05)       & $-18.60(0.30)$ & $-18.73(0.28)$ &                             &          & 0.06(0.03)   & 32.82(0.25) & E2          & J06, VSNET         \\
1992bo &       1.69(0.05)       & $-18.61(0.17)$ & $-18.59(0.15)$ &                             &          & 0.03(0.03)   & 34.34(0.12) & SB0$_\mathrm{pec}$ & H96a, H96b, P99     \\
1993H  &       1.70(0.10)       & $-18.57(0.25)$ & $-18.73(0.20)$ &         0.50(0.05)          &   73(8)  & 0.12(0.06)   & 35.07(0.09) & SBab(rs)    & H96a, H96b, P99     \\
1986G  &       1.81(0.07)       & $-17.76(0.32)$ & $-18.03(0.26)$ &         0.60(0.04)          &   68(4)  & 0.78(0.07)   & 27.61(0.11) & S0$_\mathrm{pec}$  & H87, P87, P99,     \\
       &                        &                &                &                             &       & $R_V\!\!=\!2.4^e$ &           &             & F00, B05        \\
1998bp &       1.83(0.06)       & $-17.73(0.25)$ & $-18.08(0.24)$ &                             &          & 0.08(0.03)   & 33.13(0.22) & E           & G04, J06               \\
1997cn &       1.88(0.10)       & $-17.17(0.20)$ & $-17.79(0.19)$ &         0.68(0.07)          &   75(9)  & 0.03(0.03)   & 34.25(0.13) & E           & T98, B05, J06      \\
1999by &       1.90(0.05)       & $-17.17(0.26)$ & $-17.65(0.25)$ &         0.69(0.05)          &   97(4)  & 0.02(0.03)   & 30.75(0.23) & Sb          & T00, H01, V01,     \\
       &                        &                &                &                             &          &              &             &             & G04, VSNET  \\
2005bl &       1.93(0.10)       & $-17.24(0.34)$ & $-17.85(0.27)$ &         0.63(0.06)          &   99(9)  & 0.20(0.08)   & 35.10(0.09) & E           & this work          \\
1991bg &       1.94(0.10)       & $-16.85(0.34)$ & $-17.56(0.29)$ &         0.66(0.05)          &  106(5)  & 0.08(0.06)   & 31.28(0.20) & E1          & F92, L93, T96,     \\
       &                        &                &                &                             &          &              &             &             & P99, F00, T01   \\
1999da &       1.95(0.10)       & $-16.98(0.22)$ & $-17.65(0.20)$ &                             &          & 0.06(0.03)   & 33.58(0.17) & SA0         & K01                    \\
1998de &       1.95(0.09)       & $-16.74(0.19)$ & $-17.43(0.17)$ &         0.69(0.04)          &  146(3)  & 0.06(0.03)   & 34.06(0.14) & S0          & M01, J06               \\
\hline
\end{tabular}
\\[1.5ex]
$^a$~Post-maximum decrease of the Si\,{\sc ii} $\lambda6355$ velocity in km\,s$^{-1}$\,d$^{-1}$, 
see \citet{Benetti05} ans Section~\ref{Spectroscopic parameters}.\quad
$^b$~Total (Galactic + host galaxy) colour excess.\quad
$^c$~Distance modulus from Cepheids (SNe~1986G and 1999by)\footnotemark, SBF and PNLF 
(SN~1991bg) or the host galaxy recession velocity with respect to the CMB rest frame (NED) assuming 
$H_0=72\,\rmn{km}\,\rmn{s}^{-1}\rmn{Mpc}^{-1}$ (other~SNe). For the latter, an uncertainty 
of $300$\,km\,s$^{-1}$ has been adopted to account for the galaxies' peculiar motions.\quad
$^d$~Host-galaxy morphology (from LEDA).\quad
$^e$~Spectropolarimetry \citep{Hough87} indicates $R_V = 2.4 \pm 0.13$ for the dust in Centaurus A, the host of SN~1986G. 
We adopted this value and propagated the assigned error to the uncertainty in the absolute magnitudes.\\[1.9ex] 
H96a = \citet{Hamuy96a};\quad
H96b = \citet{Hamuy96b};\quad
P99 = \citet{Phillips99};\quad
J06 = \citet{Jha06};\quad
H87 = \citet{Hough87};\quad
P87 = \citet{Phillips87};\quad
F00 = \citet{Ferrarese00};\quad
B05 = \citet{Benetti05};\quad
G04 = \citet{Garnavich04};\quad
T98 = \citet{Turatto98};\quad
T00 = \citet{Toth00};\quad
H01 = \citet{Howell01};\quad
V01 = \citet{Vinko01};\quad
F92 = \citet{Filippenko92b};\quad
L93 = \citet{Leibundgut93};\quad
T96 = \citet{Turatto96};\quad
T01 = \citet{Tonry01};\quad
K01 = \citet{Krisciunas01};\quad
M01 = \citet{Modjaz01};\quad
VSNET = Variable Star Network
\end{footnotesize}
\end{table*}
\footnotetext{For SN~1986G the Cepheid measurement does not refer to the actual 
host galaxy Centaurus A, but to NGC 5253, which is a member of the same group 
\citep{Ferrarese00}. It was adopted here for the lack of more reliable distance 
estimates.}

\subsection{Photometric behaviour of underluminous SNe~Ia}
\label{Photometric behaviour of underluminous SNe Ia}

Here we focus on the photometric properties of underluminous SNe~Ia, 
compare their light-curve shapes and colour indices, determine their peak 
absolute magnitudes, and investigate their decline-rate vs. luminosity 
relationship.

\subsection*{Light- and colour-curve morphology}

\begin{figure}   
   \centering
   \includegraphics[width=8.4cm]{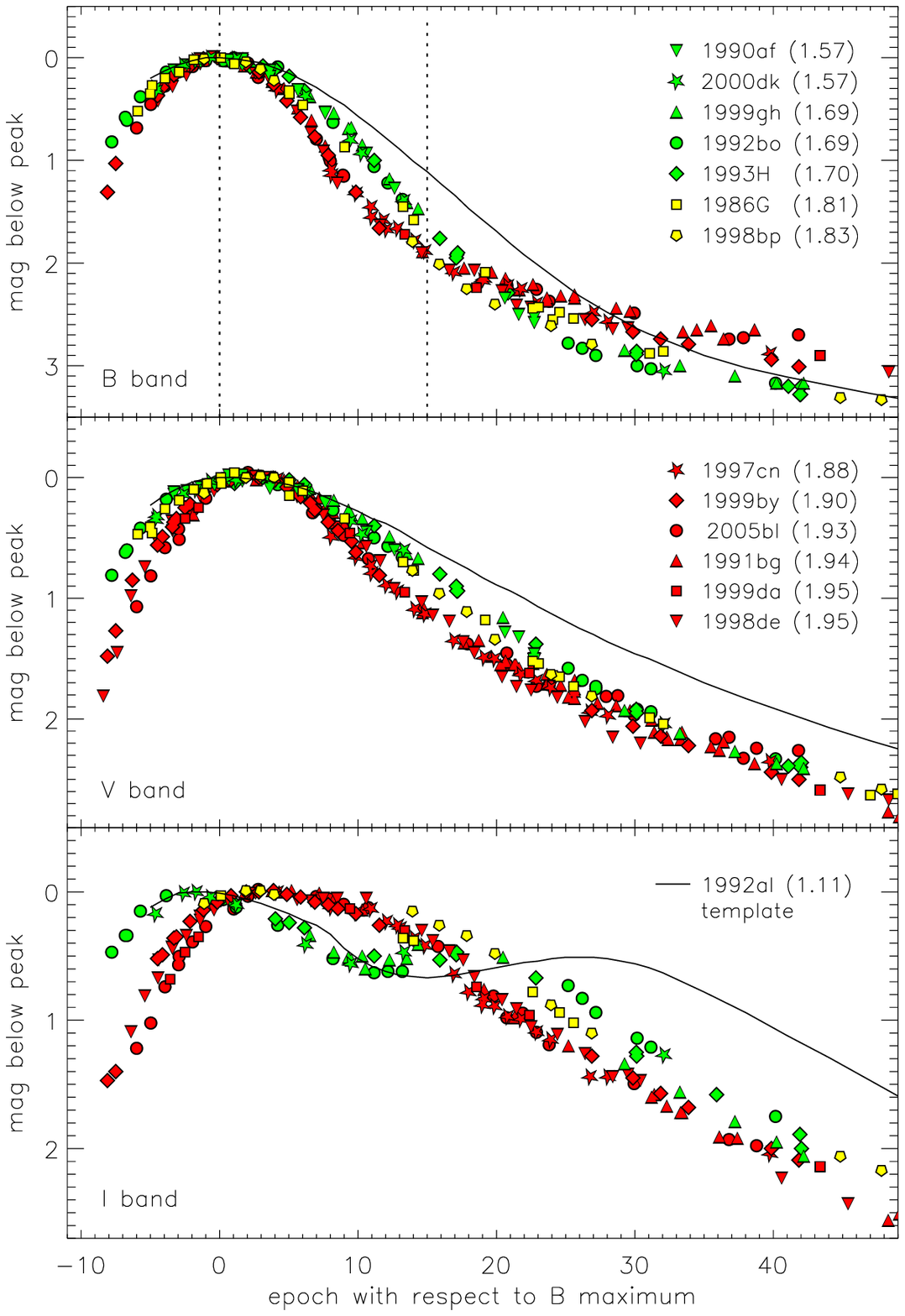}
   \caption{$BV\!I$ light curves of the rapidly-declining SNe~Ia of 
   Table~\ref{SNe_faint}, normalised to their peak magnitudes. The 
   $\Delta$m$_{15}(B)_\mathrm{true}$ of each SN is given in parentheses. 
   Different symbol colours represent different decline rates
   (green: $1.50<\Delta$m$_{15}(B)_\mathrm{true}<1.75$; 
   yellow: $1.75\leq \Delta$m$_{15}(B)_\mathrm{true}<1.85$); 
   red: $\Delta$m$_{15}(B)_\mathrm{true}\geq 1.85$). 
   Light-curve templates of SN~1992al ($\Delta$m$_{15}(B)_\mathrm{true} 
   = 1.11$, \citealt{Hamuy96c}) are shown for comparison.}
   \label{fig:LC_faint}
\end{figure}

Fig.~\ref{fig:LC_faint} compares the $B$, $V$ and $I$ light curves of all the 
objects in Table~\ref{SNe_faint}, rescaled to coincide at maximum light.

The SNe with $\Delta$m$_{15}(B)_\mathrm{true}$\,$\sim$\,$1.9$ (red symbols in 
Fig.~\ref{fig:LC_faint}), including SN~2005bl, show a single-peaked broad 
$I$-band light curve, whose maximum is delayed by a few days with respect to 
that in $B$. In contrast, the SNe with $\Delta$m$_{15}(B)_\mathrm{true}$\,$
\sim$\,$1.5$ to $1.75$ (green symbols) clearly have double-peaked $I$-band 
light curves, and, as in normal-luminosity SNe~Ia, the first $I$-band peak 
precedes that of the $B$ band \citep[e.g.][]{Hamuy96c,Leibundgut00}. These 
observations are in agreement with the results of \citet{Kasen06}, who computed 
synthetic light curves with the \textsc{sedona} code, varying the mass of 
$^{56}$Ni. His prediction is an ever smaller temporal offset of the first and 
secondary $I$-band maxima with decreasing SN luminosity, and the merging to a 
single broad peak for the most underluminous objects. The $B$-band light curves 
of the two groups (red and green symbols in Fig.~\ref{fig:LC_faint}) also 
exhibit noticeable differences in shape, probably influenced by the evolution 
of the Ti trough. 91bg-like SNe settle to the radioactive tail earlier, only 
$\sim$\,$15$\,d after maximum light. Consequently, the drop in magnitude from 
the peak to the tail is smaller for these, although their initial decline is 
steeper. Thus, the $B$ light curves of 91bg-like and ``normal'' SNe~Ia cannot 
be transformed into each other by employing a simple strech factor.
 
The only objects in the sample with $1.75 \leq \Delta$m$_{15}(B)_\mathrm{true} 
< 1.85$ are SNe~1986G and 1998bp (yellow symbols). The light-curve morphology 
of these SNe seems to provide a link between the formerly defined groups. While 
the $I$ band is still single-peaked with maybe a hint of a double peak, the 
luminosity drop from maximum to the radioactive tail in the $B$ band resembles 
that of SNe with $\Delta$m$_{15}(B)_\mathrm{true} < 1.75$. The existence of 
objects with such intermediate properties might support the idea of a common 
progenitor and explosion scenario for \textit{all} SNe~Ia. However, SN statistics 
tell us that objects with $\Delta$m$_{15}(B)_\mathrm{true} \approx 1.8$ are 
intrinsically rare. 

\begin{figure}   
   \centering
   \includegraphics[width=8.4cm]{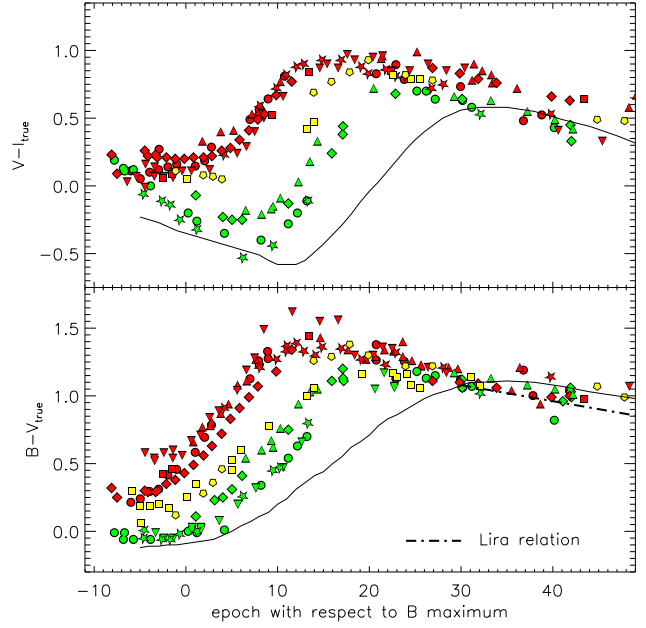}
   \caption{Evolution of the $B-V$ (top panel) and $V-I$ (bottom panel) 
   colour indices of the rapidly-declining SN~Ia sample of Table~\ref{SNe_faint} 
   plus SN~1992al. The symbols and colour coding are the same as in Fig.~\ref{fig:LC_faint}. 
   The dash-dotted line in the lower panel represents the \citet{Lira95} relation, 
   i.e. the uniform $B-V$ colour that all SNe~Ia are supposed to exhibit between 
   day $+30$ and $+90$.}
   \label{fig:colours_faint}
\end{figure}

Fig.~\ref{fig:colours_faint} shows the reddening-corrected $B-V$ and $V-I$ 
colour curves of the same SNe as above. The differences in $B-V$ after $+30$\,d 
are marginal, and all SNe obey the \citet{Lira95} relation within the 
uncertainties (note, however, that for the more strongly-reddened SNe~1986G and 
2005bl the Lira relation was directly or indirectly used to infer the true 
extinction along the line of sight, making this a circular argument). On the 
contrary, the $B-V$ colour evolution around maximum light shows remarkable 
differences, the 91bg-like SNe~Ia having a much redder colour at maximum ($0.4$ to 
$0.7$ mag, compared to $\sim$\,$0.0$ mag for SNe~Ia with $\Delta$m$_{15}(B)_\mathrm{true} 
< 1.75$). Also, the peak $B-V$ colour is reached earlier in underluminous SNe, 
and is redder by $\sim$\,$0.3$ mag.

In $V-I$ the differences among the objects of Table~\ref{SNe_faint} are even more 
pronounced.The $V-I$ colour index of SNe~Ia with $1.50 < \Delta$m$_{15}(B)_\mathrm{true} 
< 1.75$ decreases from about $10$\,d before to $5$--$10$\,d after $B$-band maximum just 
like in all intermediate or slow decliners, then increases steeply, levelling off at 
$\sim$\,$20$\,d and again decreasing slowly thereafter. 91bg-like SNe, in contrast, 
do not show the initial bluening, but become redder from the very first available 
observations onwards. This behaviour continues until $\sim$\,$15$\,d past maximum, 
when the curves flatten and subsequently a soft bluening sets in. Evidently, the 
origin of the different $V-I$ evolution lies in the different delay of the main 
maximum and the presence or absence of the secondary maximum in the $I$ band.

\subsection*{Absolute magnitudes \& Phillips relation}

\begin{figure}   
   \centering
   \includegraphics[width=8.4cm]{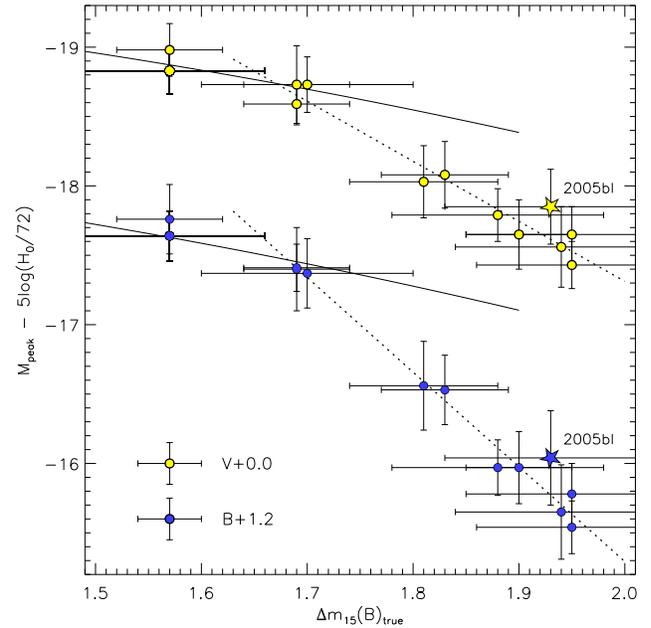}
   \caption{Peak absolute $B$- and $V$-band magnitudes of the SN sample of 
   Table~\ref{SNe_faint}. The dotted lines are the best linear fits to the data 
   with $\Delta$m$_{15}(B)_\mathrm{true} \geq 1.69$, characterised by slopes of 
   $6.83 \pm 0.32$ and $4.33 \pm 0.31$ for $B$ and $V$, respectively. The solid 
   lines show the quadratic relation obtained by \citet{Phillips99} for SNe~Ia 
   with $\Delta$m$_{15}(B)_\mathrm{true} < 1.70$.}
   \label{fig:absolute_mags}
\end{figure}

Table~\ref{SNe_faint} shows that the peak absolute magnitudes of the SNe with 
$\Delta$m$_{15}(B)_\mathrm{true} > 1.5$ span a wide range, from $-16.74$ to 
$-18.96$ in the $B$ band. As \citet{Hamuy96a}, \citet{Phillips99} and 
\citet{Garnavich04} pointed out, also for these SNe the peak magnitudes 
correlate with $\Delta$m$_{15}(B)_\mathrm{true}$, but the dependence is 
steeper than for more slowly declining SNe~Ia. Based only on SNe with 
$\Delta$m$_{15}(B)_\mathrm{true} \geq 1.69$, our best linear fits 
(Fig.~\ref{fig:absolute_mags}) are given by

\begin{small}
\begin{displaymath}
M_{B,\mathrm{peak}} = -18.54 + 5\log(H_0/72) + 6.83\,(\Delta m_{15}(B)_\mathrm{true}-1.7) 
\end{displaymath}

\begin{displaymath}
M_{V,\mathrm{peak}} = -18.61 + 5\log(H_0/72) + 4.33\,(\Delta m_{15}(B)_\mathrm{true}-1.7). 
\end{displaymath}
\end{small}
The slopes of $6.83 \pm 0.32$ and $4.33 \pm 0.31$ for the $B$ and $V$ bands are 
in excellent agreement with the results of \citet{Garnavich04}, and much steeper 
than what \citet{Hamuy96a} find for SNe with $\Delta$m$_{15}(B)_\mathrm{true} 
\leq 1.69$ ($0.78 \pm 0.17$ and $0.71 \pm 0.14$ for $B$ and $V$, respectively). 
Also the quadratic relation derived by \citet{Phillips99} for SNe with 
$\Delta$m$_{15}(B)_\mathrm{true} < 1.70$ provides a poor fit to the peak 
magnitudes of fast decliners.

\subsection{The rise time of underluminous SN\lowercase{e}~I\lowercase{a}}
\label{Rise time}

Rise times of SNe~Ia are important to constrain possible explosion models. In 
our synthetic spectra the rise time $t_\mathrm{r}$ determines the actual 
(scaled) density profile at a given epoch (cf. Section~\ref{Code concept}), and 
hence the matter density of the line-forming region. Here we report our attempts 
to estimate the rise time of 91bg-like SNe, in particular of SN~2005bl.

\subsection*{Photometric rise-time determination}

Efforts have been made to measure SN-Ia rise times directly from a fit to 
their early-time light curves \citep{Riess99,Conley06,Strovink07}. Depending 
on measurement details, these studies yield rise times of $17$ to $20$\,d for 
fiducial $\Delta$m$_{15}(B)_\mathrm{true}=1.1$ SNe~Ia. Furthermore, there 
seems to be a negative correlation between $t_\mathrm{r}$ and 
$\Delta$m$_{15}(B)_\mathrm{true}$, faster decliners having shorter rise times. 
However, all these studies are based on samples of ``normal'' SNe~Ia with 
$\Delta$m$_{15}(B)_\mathrm{true}\leq1.5$, and it may be doubted whether the 
inferred trends can be extrapolated to 91bg-like SNe with 
$\Delta$m$_{15}(B)_\mathrm{true} \approx 1.9$.

\begin{figure}   
   \centering
   \includegraphics[width=8.4cm]{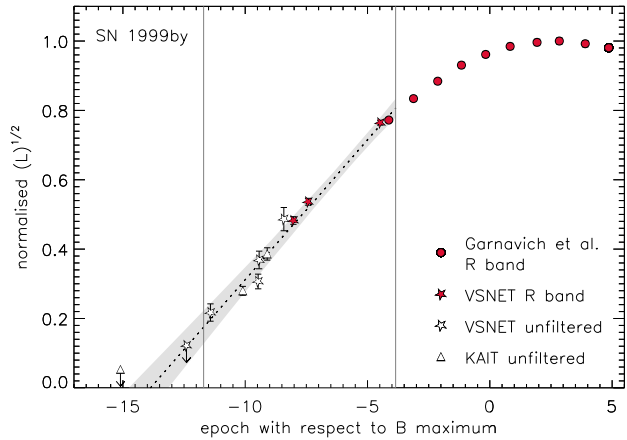}
   \caption{R-band\,/\,unfiltered early-time light curve of SN~1999by, plotted as 
   $L^{1/2}$ vs. $t-t_{B_\mathrm{max}}$, and linear fit (dotted line) to the data 
   up to day $-4$ (between the two thin vertical lines). Assuming $L\propto t^2$, 
   the instance of explosion would be given by the intersection of the fit with 
   $L^{1/2}=0$ at $-13.9$\,d. The grey-shaded region marks the 3-$\sigma$ 
   confidence bands of the linear fit. Caveats are discussed in the text.}
   \label{fig:99by_rise}
\end{figure}

What makes these direct measurements so demanding is the need for high-quality 
early-time photometry, preferentially before day $-10$ \citep{Riess99}, which 
to date is not available for any 91bg-like SN, including 2005bl. In order to 
constrain the rise times of this class of objects, it may be useful to consider 
the two 91bg-like SNe~Ia with the earliest photometric data, SNe~1998de 
and 1999by. Both have a filtered light-curve coverage starting about one week 
before $B$-band maximum, complemented by earlier unfiltered measurements 
and some deep detection limits shortly before. For SN~1998de the earliest 
detection showing just ``a hint of the supernova'' was on unfiltered CCD 
frames taken $12.3$\,d before $B$-band maximum (preceded by a non-detection 
down to a limiting magnitude of $19.0$ five days earlier, \citealt{IAUC6977}), 
thus providing at least a lower limit for the rise time. 
For SN~1999by, the earliest detection on unfiltered CCD frames dates back to 
day $-11.4$, showing the SN at only $5\pm1$\,\% of the peak $R$-luminosity 
\citep{IAUC7156}. Further detection limits constrain the SN luminosity to be 
less than $\sim$\,$1.6$\,\% of the peak value at $-12.4$\,d, and less than 
$\sim$\,$0.3$\,\% at $-15.1$\,d. Applying the method described by \citet{Riess99} 
to these data, but using unfiltered and $R$-band points between $-11.4$ and 
$-4.0$\,d for the fit (Fig.~\ref{fig:99by_rise}), a rise time to $B$-band 
maximum of $13.9_{-1.1}^{+1.2}$\,\d would be inferred for SN~1999by, the 
errors being the 3-$\sigma$ confidence levels of the fit combined with the 
uncertainty in determining the epoch of $B$-band maximum \citep{Garnavich04}. 

In this approach we disregarded a possible (and theoretically not unexpected) 
deviation from the early $L\propto t^2$-behaviour which is the basis of the 
Riess et~al. analysis. If the synthesised $^{56}$Ni is confined to the inner 
ejecta, which -- at least for 91bg-like objects -- is supported by the lack of 
Ni in the $-6$ and $-5$\,d spectra of SN~2005bl (cf. Section~\ref{Parameters 
and composition}), it should require some time for the photons to reach the 
photosphere and be released. This could result in an initial post-explosion 
phase with only very little brightening of the SN, before the observed steep 
rise of the light curve sets in. For this reason, the inferred value of $13.9$\,d 
might be considered a \textit{lower limit} to the actual rise time of SN~1999by. 
Given the similarity in terms of light-curve shape, this result should provide 
a good indication also for the rise time of SN~2005bl.

\subsection*{The rise time addressed by spectral modelling}

The rise time constitutes an important input parameter for our spectrum synthesis 
calculations. Since the colour of an observed spectrum has to be reproduced 
in a corresponding model, the photospheric temperature, which can be crudely 
approximated through the Stefan-Boltzmann-law $L \propto v_\mathrm{ph}^2 t^2 T^4$,
is confined to a limited range. Assuming a shorter rise time, the photospheric 
velocity thus usually has to be increased in order to keep the photospheric 
temperature constant. The line velocities in the synthetic spectrum then also 
tend to be larger.

Given the difficulties in observational rise-time determination, we created 
eight models with different rise times ($12.5$--$25.0$\,d) for the earliest 
available spectrum of SN~2005bl ($-6$\,d), where a change in $t_\mathrm{r}$ 
has the largest relative effect. Excluding models largely deviating from the 
observed spectrum yielded a range of acceptable rise times. For the models 
with $t_\mathrm{r}<14.0$\,d, we had to use photospheric velocities of 
$v_\mathrm{ph}>10\,200$ km\,s$^{-1}$, which are incompatible with observed and 
measured (Fig.~\ref{fig:ejecta_velocities}) O\,{\sc i} and S\,{\sc ii} expansion 
velocities. Choosing rise times of $14.0$--$15.5$\,d, most weaker lines were 
fitted nicely, while in stronger lines such as Si\,{\sc ii} $\lambda 6355$ or 
Ca\,{\sc ii} H\&K there was still a lack of absorption at low velocity. While 
weaker lines mostly form near the photosphere, strong lines have considerable 
strength also in layers well above. Thus, their centroid typically shows a 
larger blueshift, and their emission, centred at the rest wavelength, is more 
pronounced. In order to decrease the flux in their red wings, $v_\mathrm{ph}$ 
had to be reduced to $5000$--$6000$ km\,s$^{-1}$, corresponding to rise times 
of $22$--$20$\,d. The weaker features, however, then appeared at velocities 
too low in the synthetic spectrum. 

The difficulty to match all line-velocities simultaneously points to some 
substantial shortcoming in our treatment of the SN ejecta. Stratifying the 
ejecta composition \citep{Stehle05} assuming only little Si and Ca at high 
velocity might cure some of the problems, as it could potentially remove 
the blue parts of the respective strong absorption lines and reduce their 
re-emission. Yet, we retained our uniform-abundance approach, because no 
observational data earlier than $-6$\,d are available to constrain the 
additional parameters of a stratified-abundance model. Alternative 
explanations for the inconsistencies include a possible inadequacy of 
the W7 density structure, a deviation of the total ejected mass from 
$M_\mathrm{Ch}$, or 3D effects. 

Nonetheless, based on our model sequence, we confidently exclude rise times 
shorter than $14$ and longer than $22$\,d. For all spectral models discussed 
below we assumed a rise time of $17.0$\,d. This value is fully compatible with 
the lower limits observationally derived above. At the same time, it is slightly 
shorter than what is favoured for $\Delta$m$_{15}(B)_\mathrm{true}=1.1$-SNe, 
in agreement with the trend found for SNe~Ia with $\Delta$m$_{15}(B)_\mathrm{true} 
\leq 1.5$ \citep[e.g.][]{Riess99,Kasen07}.

\subsection{Spectroscopic parameters}
\label{Spectroscopic parameters}

\begin{figure}   
   \centering
   \includegraphics[width=8.4cm]{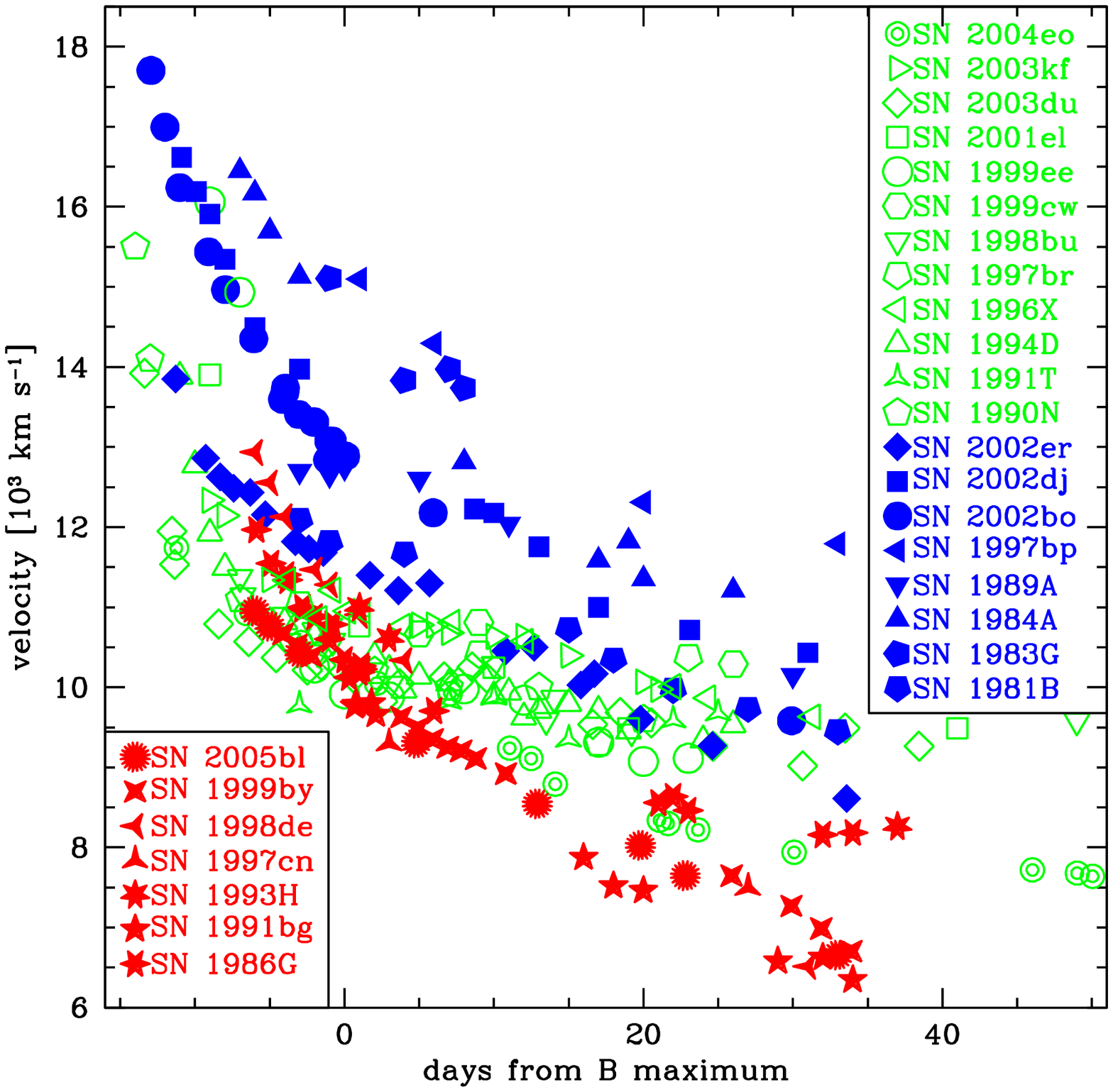}
   \caption{Velocity measured from the minimum of the Si\,{\sc ii} $\lambda6355$ 
   line as a function of time. Open green symbols represent low-velocity-gradient 
   (LVG) SNe, filled blue symbols high-velocity-gradient (HVG) SNe, and red starred 
   symbols correspond to the FAINT SNe, as defined by \citet{Benetti05}. The values 
   for HVG and LVG SNe are taken from Benetti et al. and \citet{Pastorello07b}, 
   while the data set for the FAINT class has been extended and in parts remeasured.}
   \label{fig:velocities}
\end{figure}

\begin{figure}   
   \centering
   \includegraphics[width=8.4cm]{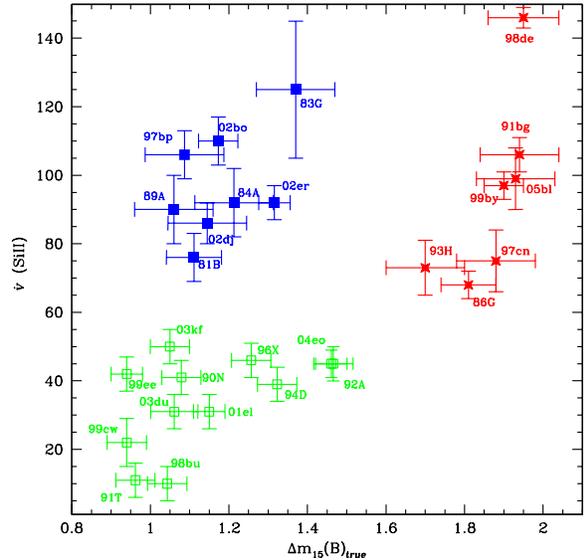}
   \caption{Si\,{\sc ii} $\lambda6355$ post-maximum velocity gradient $\dot v$ 
   \citep{Benetti05} versus $\Delta$m$_{15}(B)_\mathrm{true}$. Filled blue squares 
   are HVG SNe, open green squares LVG SNe, and filled red stars FAINT SNe.} 
   \label{fig:dm15_vdot}
\end{figure}

\begin{figure}   
   \centering
   \includegraphics[width=8.4cm]{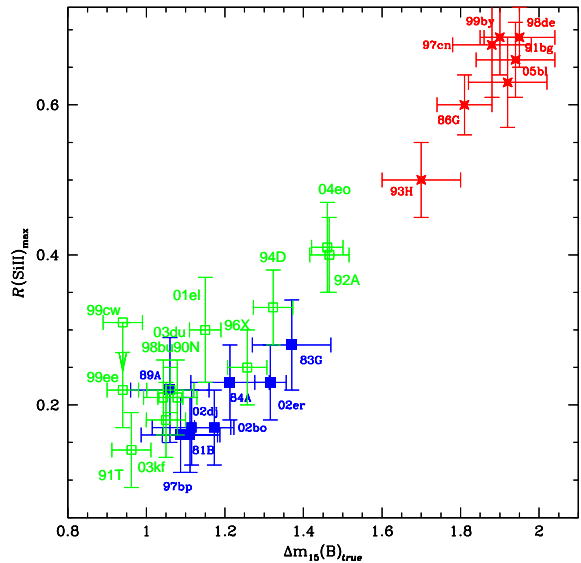}
   \caption{Si line-depth ratio $\mathcal{R}$(Si)$_\mathrm{max}$ \citep{Nugent95} 
   as a function of SN decline rate $\Delta$m$_{15}(B)_\mathrm{true}$. The 
   strong correlation between these two parameters is evident. Symbols as in 
   Fig.~\ref{fig:dm15_vdot}.}
   \label{fig:dm15_R(Si)}
\end{figure}

\begin{figure}   
   \centering
   \includegraphics[width=8.4cm]{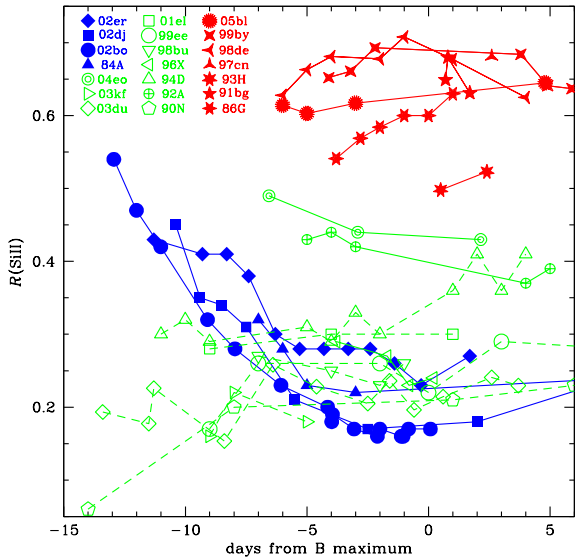}
   \caption{Time-evolution of the \citet{Nugent95} ratio $\mathcal{R}$(Si) before 
   and near maximum light. The same symbols as in Fig.~\ref{fig:velocities} have 
   been used. Interpolating lines between adjacent points have been drawn to 
   guide the eye.}
   \label{fig:t_R(Si)}
\end{figure}

A spectroscopic comparison of SN~2005bl with other underluminous SNe~Ia 
was performed in Section~\ref{Spectroscopic comparison}, in particular in 
Figs.~\ref{fig:comp_minus05}--\ref{fig:comp_plus33}. Here, we concentrate 
on parameters such as $\mathcal{R}$(Si) \citep{Nugent95} and $\dot v$ (the 
average daily rate of velocity decrease of Si\,{\sc ii} $\lambda6355$ between 
$B$-band maximum and either the time the Si\,{\sc ii} feature disappears or 
the last available spectrum, whichever is earlier; \citealt{Benetti05}). With 
respect to \citet{Benetti05}, these parameters have been remeasured for some 
FAINT SNe using additional data, and SNe~1998de \citep{Matheson07} and 2005bl 
have been added.

Fig.~\ref{fig:velocities} shows the velocity evolution of the Si\,{\sc ii} 
$\lambda6355$ line of SN~2005bl and a large set of comparison objects. Once 
more, SN~2005bl is extremely similar to SNe~1991bg, 1997cn and 1999by. Before 
maximum light, the Si velocities of these SNe are comparable to those of the 
slower members of Benetti et al.'s LVG group, but the velocity decrease is 
much faster. Hence, after maximum light, they have clearly the lowest expansion 
velocities. SN~1998de, photometrically nearly a twin of SN~2005bl, has 
noticeably higher expansion velocities at early phases. Whether this is caused 
by a high-velocity component \citep{Mazzali05}, by differences in the density 
structure, or by a higher kinetic energy is difficult to decide without detailed 
modelling. High-velocity features could provide a natural explanation for the 
observed differences without the need to change the explosion energetics. On the 
other hand the Si\,{\sc ii} $\lambda6355$ line of SN~1998de looks symmetric, and 
the velocities are higher than in other 91bg-like SNe not only at the earliest 
phases, but also near maximum light, when in most other SNe the high-velocity 
features have disappeared \citep{Mazzali05,Garavini07}. 
Finally, the transitional SNe~1993H and 1986G have higher velocities than 
SN~2005bl and the other 91bg-like SNe (except SN~1998de), more similar to the 
LVG group. Their comparatively shallow post-maximum velocity decrease $\dot v$ 
\citep{Benetti05} also shows their proximity to LVG SNe.

In the $\dot v$ vs. $\Delta$m$_{15}(B)_\mathrm{true}$ plane, FAINT SNe seem to 
form a separate cluster (Fig.~\ref{fig:dm15_vdot}). However, this impression 
may be caused by the lack of SNe with $\Delta$m$_{15}(B)_\mathrm{true}$ between 
$1.5$ and $1.7$ in the Benetti et al. sample. Whether or not this gap is real 
is not easy to decide given the poor statistics for rapidly-declining SNe~Ia. 
With respect to $\dot v$ alone, FAINT SNe are very heterogeneous, with values 
ranging from those typical of LVG SNe to the highest values ever recorded in 
a SN~Ia (for SN~1998de). The mean $\dot v$ of $95 \pm 27$ (statistical) 
km\,s$^{-1}$\,d$^{-1}$ is similar to that reported by Benetti et~al. for HVG 
SNe ($97 \pm 16$ km\,s$^{-1}$\,d$^{-1}$). Within the FAINT group there may be 
a tendency for higher $\dot v$ with larger $\Delta$m$_{15}(B)_\mathrm{true}$, 
but there are too few objects and their range in $\Delta$m$_{15}(B)_\mathrm{true}$ 
is too small to postulate a correlation. 

Concerning $\mathcal{R}$(Si)$_\mathrm{max}$, the value of $\mathcal{R}$(Si) at 
$B$-band maximum, we confirm the observed trend of larger values for faster 
decliners (Fig.~\ref{fig:dm15_R(Si)}). The origin of this behaviour was the 
subject of a number of studies, as an explanation is not straightforward. 
\citet*{Hachinger06} have shown that the change in $\mathcal{R}$(Si)$_\mathrm{max}$ 
with $\Delta$m$_{15}(B)_\mathrm{true}$ is almost solely caused by a variation 
in the strength of the Si\,{\sc ii} $\lambda5972$ line, which is somewhat 
counter-intuitive as the excitation of this line should be favoured by higher 
temperatures. \citet{Garnavich04} suggested that in underluminous SNe~Ia the 
feature at $\sim$\,$5800$\,\AA\ might be dominated by Ti lines. However, from 
the synthetic spectra shown in Section~\ref{Parameters and composition} we cannot 
confirm this option. Instead, \citet[][and in prep.]{Hachinger07} demonstrate 
that the observed tendency is a combined ionisation and excitation effect, 
natural rather than unexpected.

The pre-maximum time-evolution of $\mathcal{R}$(Si), which was also used by 
\citet{Benetti05} to distinguish between the LVG and HVG groups, cannot be 
studied equally well for the FAINT SNe because of the lack of sufficiently early 
data (more than 5\,d before maximum). In fact, only before day $-5$ do LVG 
and HVG SNe show significant differences in their behaviour, $\mathcal{R}$(Si) 
decreasing with time for HVG SNe, and being constant or increasing slightly 
for LVG SNe. Near maximum light, the evolution seems to be fairly flat for all 
SNe~Ia, including the FAINT SNe which merely exhibit larger absolute values of 
$\mathcal{R}$(Si) (see Fig.~\ref{fig:t_R(Si)}). 
However, especially the Si\,{\sc ii} $\lambda5972$ line may suffer from variable 
contributions of other elements, notably Na\,{\sc i}\,D at later phases when the 
spectra become cooler. This limits the conclusive power of all studies related 
to the evolution of $\mathcal{R}$(Si) after maximum light, as already at day 
$+5$ Na\,{\sc i}\,D is visible as a shoulder in the blue wing of the Si\,{\sc ii} 
$\lambda5972$ feature (cf. Section~\ref{Spectra of SN 2005bl}).

\subsection{Element abundances from synthetic spectra} 
\label{Element abundances from synthetic spectra} 

Spectrum synthesis calculations offer an insight into the chemical 
composition of the ejecta, making it possible to trace the nuclear reactions 
that took place during explosive burning. However, because of the limitations 
and possible shortcomings of the models mentioned in Section~\ref{Spectral 
modelling} (i.e., no abundance stratification, use of a Chandrasekhar-mass 
progenitor and of the W7 density distribution), the exact numbers in 
Table~\ref{model parameters} should be taken with caution. Therefore, we 
confine ourselves to analysing clear trends, which should be robust with 
respect to refined models.

\subsection*{Unburned material, carbon detection}

As for all rapidly-declining SNe~Ia the spectrum of SN~2005bl is characterised 
by strong O\,{\sc i}\,$\lambda7774$. The strength of this line is largely a 
temperature effect, as the comparatively low temperatures encountered in 
91bg-like SNe~Ia result in a fair fraction of neutral oxygen besides the singly 
ionised state which normally dominates the ejecta. Nevertheless, it also 
signals the presence of a significant amount of unburned material in the SN 
ejecta.\footnote{In this section we refer to C and O as unburned, although -- 
depending on the initial C-to-O ratio of the white dwarf -- a significant 
fraction of the O may have been produced during the explosion by incomplete C 
burning.} This is fully consistent with the often-anticipated low burning 
efficiency of these objects (from light curves and nebular spectra one can 
infer that the amount of synthesised $^{56}$Ni is $\sim$\,$0.1$\,$M_\odot$, 
which is at most a quarter of that in normal-luminosity SNe~Ia, 
\citealt{Stritzinger06}). Direct information on the oxygen abundance is 
difficult to derive from synthetic spectra, since O\,{\sc i}\,$\lambda7774$ 
is the only strong feature of this element at optical wavelengths, and is 
heavily saturated in all our models. Hence, moderate changes in the oxygen 
content have no impact on the strength of this feature. Additionally, our 
abundance estimate should be considered an upper limit, as, guided by nuclear 
reaction network calculations, we ascribe the entire mass without observable 
signatures to oxygen.

\begin{figure}   
   \centering
   \includegraphics[width=8.4cm]{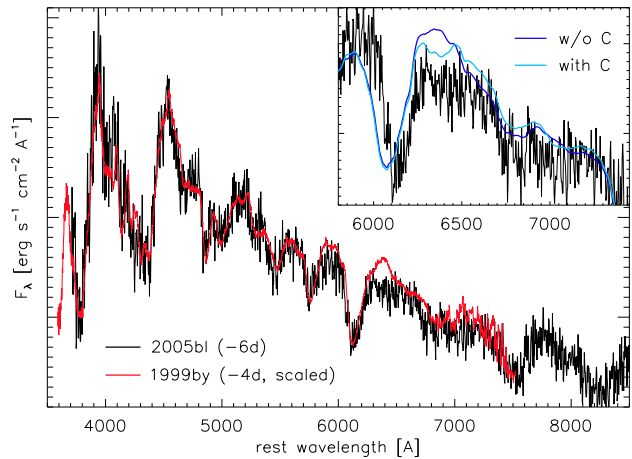}
   \caption{Identification of carbon in the spectra of SN~2005bl. A comparison 
   of the $-6$\,d spectrum of SN~2005bl and the $-4$\,d spectrum of SN~1999by 
   shows a flux deficit at $6400$\,\AA\ in SN~2005bl, likely caused by C\,{\sc 
   ii} $\lambda6580$. The insert shows synthetic spectra for day $-6$, one with 
   the composition reported in Table~\ref{model parameters} including $4.5$\% 
   of C, the other without any C.}
   \label{fig:carbon}
\end{figure}
 
Applying this strategy, we find oxygen mass fractions ranging from $> 90\%$ at 
day $-6$ to $\sim$\,$80\%$ at day $+5$. With a W7 density profile and 
photospheric velocities of $7500$ and $6000$ km\,s$^{-1}$, about $50$ and 
$60\%$ of the ejecta mass are located above the photospheres at days $-6$ and 
$+5$, respectively. Hence, if a Chandrasekhar-mass explosion with W7 density 
profile provides an acceptable model for 91bg-like SNe~Ia, at least the outer 
$50\%$ of the ejecta of SN~2005bl and other members of this group are entirely 
dominated by unburned material. These numbers are significantly different from 
the results obtained by modelling normal SNe~Ia \citep[cf., e.g.,][]{Stehle05}. 
They are also in disagreement with the picture sketched in the Zorro plot 
\citep{Mazzali07}, i.e. that all LVG and FAINT SNe~Ia have about the same amount 
of unburned material, and that the observed spectrophotometric sequence among 
SNe~Ia has its origin only in a variable ratio of $^{56}$Ni to IMEs. 

Another indication of the low burning efficiency in SN~2005bl is the likely 
detection of unprocessed carbon in its early spectra. Although the S/N of the 
$-6$ and $-5$\,d spectra is low, a clear flux deficit redwards of the Si\,{\sc ii} 
$\lambda6355$ absorption can be discerned compared to coeval spectra of SN~1999by 
(Fig.~\ref{fig:carbon}). 

Synthetic spectra enable us to investigate the effect of including different 
chemical species on this region. There are only few lines which could potentially 
contribute to the observed absorption.\footnote{Conversely, because of the 
relative lack of strong lines, the region around $6400$\,\AA\ proved to be a 
preferential window for flux redistributed from shorter wavelengths to escape. 
All elements which create opacity in the blue and UV part of the spectrum 
(Fe-group elements, Ti, Cr, other metals) rather led to an increase of the 
$6400$\,\AA\ flux excess in the model. This in turn yielded constraints on the 
metal abundance.} Including hydrogen did not improve the model, since the 
resulting H$\alpha$ absorption was at too short a wavelength, and H$\alpha$ 
emission deteriorated the fit around $6500$\,\AA. 
A clearly better match to the observed spectrum was obtained introducing 
carbon, as C\,{\sc ii} lines not only cropped the flux peak at $6400$\,\AA\ 
through C\,{\sc ii} $\lambda6580$ (Fig.~\ref{fig:carbon}, insert), but also 
improved the fit in other regions, most notably around $7000$\,\AA\ (through 
C\,{\sc ii} $\lambda7234$). Therefore, we consider the presence of a few 
percent of carbon in the ejecta at days $-6$ and $-5$ very likely.

The temperature and density conditions for carbon burning are more relaxed than 
those for oxygen burning, so that carbon is burned much more completely to 
heavier elements than oxygen. In fact, only in some -- usually very early 
($\sim$\,$-10$\,d) -- optical spectra of SNe~Ia have clear signatures of carbon 
ever been found (most prominently in SN~2006gz, \citealt{Hicken07}; cf. also 
\citealt{Branch07} for an overview). None of these SNe was 91bg-like, although 
the lower burning efficiency in these objects might favour the presence of 
unprocessed carbon. The lack of spectra taken earlier than $5$\,d before 
maximum light may have so far prevented the detection of carbon. In SN~2005bl, 
however, we are confident that signatures of carbon are present less than one 
week before maximum. This constitutes one of the few detections of C\,{\sc ii} 
in SN~Ia spectra at such a relatively late epoch; previous cases include 
SNe~1996X and 2006D \citep{Thomas07}.

\subsection*{Abundance of Fe-group elements}

The Fe-group abundance of SN~2005bl of $\sim$\,$0.01\%$ at day $-6$ and 
$\sim$\,$0.4\%$ at day $+5$ is about two orders of magnitude lower than in 
normal-luminosity SNe~Ia at comparable epochs (see, e.g., \citealt{Stehle05,
Kotak05,Elias-Rosa06,Altavilla07}). At a first glance this looks perfectly 
consistent with the low luminosity and hence the low production of $^{56}$Ni 
and other NSE-elements in a 91bg-like SN~Ia. 

\begin{figure}   
   \centering
   \includegraphics[width=8.4cm]{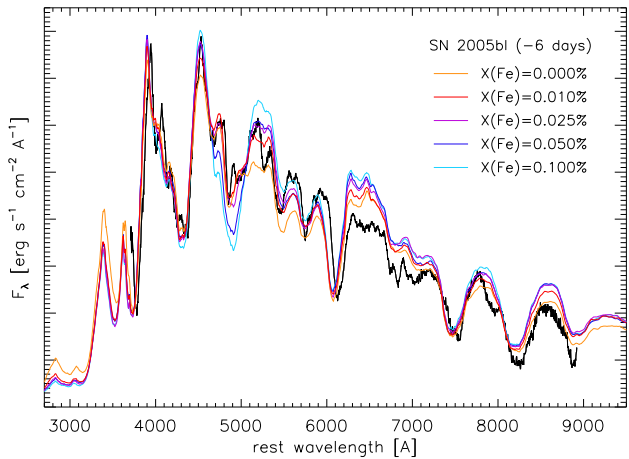}
   \caption{Sequence of synthetic spectra, varying the mass fraction of Fe 
   between $0$ to $0.1$\,\% (the latter roughly represents solar Fe abundance). 
   The observed but smoothed $-6$\,d spectrum is shown in black. The parameters 
   of the best-fitting spectrum (with X(Fe)$=0.01$\,\%) are summarised in 
   Table~\ref{model parameters}. The other spectra have been obtained by 
   changing only the amount of stable Fe at the expense of O, the most 
   abundant element, without attempting to optimise the fit by adapting other 
   parameters. Hence, better fits might be feasible also for the models with 
   X(Fe)$\neq0.01$\,\%.}
   \label{fig:X(Fe)}
\end{figure}

However, closer examination reveals that the Fe content of the pre-maximum 
models is significantly sub-solar. At $-6$\,d the best-fitting model contains 
no Ni or Co, and only $0.010$\,\% of stable Fe. Acceptable results were 
obtained with Fe abundances up to $0.025$\,\%, but also without any Fe (see 
Fig.~\ref{fig:X(Fe)}). Solar Fe abundances (the Fe mass fraction of the Sun's 
photosphere is about $0.115$\,\%, \citealt{Asplund05}) can safely be ruled out, 
as with this amount of Fe the quality of the fit deteriorates a lot around 
$5000$\,\AA. Now, the Fe content deduced not only accounts for $^{54}$Fe 
synthesised during the explosive burning (the $^{56}$Ni decay chain has too 
long decay times to give a relevant contribution at these epochs), but also 
for Fe already present in the progenitor star before the explosion. Hence, 
a sub-solar Fe content of the precursor star might be inferred.
On the other hand, SN~2005bl exploded in an elliptical galaxy, a presumably 
not metal-poor environment. If the chemical composition of the precursor did 
not deviate significantly from the average material in NGC~4070, it rather 
should have had a fairly high Fe abundance, and nuclear burning should have 
further enriched the ejecta with Fe-group elements, although this may be 
confined to the inner shells.

There may be two ways to explain the small amount of Fe required to fit the 
early-time spectra of SN~2005bl. One is to assume that Fe was not distributed 
homogeneously in the progenitor star, but concentrated towards the centre. A 
certain degree of gravitational settling of heavier elements can indeed be 
expected. Estimating whether or not this effect is sufficient to yield Fe 
abundances in accordance to those inferred from the synthetic spectra would 
require a model of gravitational diffusion inside the white dwarf, and accurate 
constraints on the mass accretion rate from the companion and the composition 
of the accreted material. Note, however, that assuming $v_\mathrm{ph} = 7500$ 
km\,s$^{-1}$ and a W7 density profile the photosphere at day $-6$ is  not 
located in the outermost layers, but half the way down the ejecta in mass 
coordinates. 
 
Alternatively, the progenitor of SN~2005bl might have had a metallicity truly 
lower than the average metallicity of an elliptical galaxy. This could be 
explained if the progenitor was rather old ($\sim$\,10 Gyr), formed before the 
ISM in NGC~4070 was metal-enriched through recurrent cycles of stellar birth, 
mass loss and death. If this scenario held true, it would provide one of the 
first direct constraints on the lifetime of an individual SN~Ia progenitor. 
Given the spectroscopic similarity of SN~2005bl to SNe~1991bg, 1997cn, 1998de 
and 1999by, and given that most of the latter SNe also exploded in early-type, 
supposedly metal-rich galaxies (see Table~\ref{SNe_faint}), this could point 
towards a very long-lived progenitor population for \textit{all} underluminous, 
91bg-like SNe~Ia. 

A speculation could be that 91bg-like SNe~Ia descend from white-dwarf mergers 
(DD scenario). White-dwarf binary systems should meet the requirement of long 
lifetimes because of the low efficiency of angular-momentum loss through 
gravitational-wave emission. In fact, in a study of the delay-time distribution 
of SNe~Ia, \citet{Greggio05} identified initially wide white-dwarf pairs as the 
only progenitor model whose delay-time distribution does not drop dramatically 
beyond $\sim$\,8 Gyr. Models of \citet{Kobayashi98} lend further support to the 
DD scenario, as they predict that no SNe~Ia should occur in SD systems for 
[Fe/H] $\leq -1$, a limit which may be exceeded by SN~2005bl. Finally, from 
spectropolarimetric observations of SN~1999by \citet{Howell01} inferred a 
degree of asphericity much higher than in ordinary SNe~Ia. On this basis they 
identified rapidly rotating white dwarfs and merging white-dwarf binaries as 
the most promising progenitor candidates for 91bg-like SNe~Ia. A different 
progenitor system with respect to ordinary SNe~Ia (for which the SD scenario 
consisting of a white dwarf and a non-degenerate companion is favoured), and 
in particular the possibility that the mass is different from $M_\mathrm{Ch}$, 
could account for observed differences in the photometric and spectroscopic 
properties of 91bg-like SNe.

\section{Conclusions}
\label{Conclusions}

We have presented and analysed optical photometric and spectroscopic data of 
the underluminous Type Ia SN~2005bl from one week before to two months after 
maximum light in $B$. In this entire interval the evolution of SN~2005bl is 
substantially different from that of slow or intermediate decliners, but very 
similar to that of the prototypical fast decliners SNe~1991bg and 1999by. At 
peak, SN~2005bl appears slightly more luminous than the latter, but the 
difference is marginal and within the error bars, which for SN~2005bl are 
dominated by uncertainties in the amount of host-galaxy extinction. With 
respect to quantities like the $\Delta$m$_{15}(B)_\mathrm{true}$ of $1.93$, the 
ejecta velocities inferred from the minimum of the Si\,{\sc ii} $\lambda6355$ 
line, and temperature indicators such as $\mathcal{R}$(Si) or the time-evolution 
of various colour indices, SN~2005bl is in fact --  within the measurement 
uncertainties -- a clone of SNe~1991bg and 1999by. 

We confirm that $\Delta$m$_{15}(B)_\mathrm{true}$ is a good luminosity indicator 
also for underluminous SNe~Ia, with a steeper dependence of the peak magnitudes 
on $\Delta$m$_{15}(B)_\mathrm{true}$ than for slow and intermediate decliners. 
From linear fits we obtain $\frac{\mathrm{d}(\mathrm{M}_\mathrm{peak})}
{\mathrm{d}(\Delta\mathrm{m}_{15}(B)_\mathrm{true})} = 6.83 \pm 0.32$ and $4.33 
\pm 0.31$ for the $B$ and $V$ band, respectively, considering objects with 
$\Delta$m$_{15}(B)_\mathrm{true} \geq 1.69$. 

We conducted an analysis analogous to that of \citet{Benetti05}, focussing on 
the behaviour of the FAINT subclass and trying to identify trends within this 
group. In addition to the SNe in Benetti et~al. we not only added the 
measurements of SN~2005bl, but also included SN~1998de \citep{Modjaz01,
Matheson07} and updated some numbers on the basis of better data availability. 
We find a correlation between $\Delta$m$_{15}(B)_\mathrm{true}$ and 
$\mathcal{R}$(Si)$_\mathrm{max}$, nicely extending the observed linear trend 
among HVG and LVG SNe to larger decline rates. Because of the lack of early 
data, the time-evolution of $\mathcal{R}$(Si) at phases where differences 
between the LVG and HVG groups become evident cannot be investigated. The 
Si\,{\sc ii} line velocities of FAINT SNe are mostly similar to those of LVG 
SNe near maximum light, but significantly lower a few weeks later. There is a 
large dispersion in post-maximum velocity gradients $\dot v$ within the FAINT 
class, with numbers comparable to those of HVG SNe, but exceeding their range 
in both directions. 

Synthetic spectra for SN~2005bl at four epochs before and soon after maximum 
light have been computed using a 1D Monte Carlo code. Despite uncertainties and 
shortcomings such as the poorly-constrained rise time, the possible inadequacy 
of the W7 density profile for 91bg-like SNe~Ia, and the fact that computations 
were performed without chemical stratification, a number of interesting results 
were obtained. The presence of carbon in the $-6$ and $-5$\,d spectra, indicated 
by a visual comparison with SN~1999by at similar epoch, was confirmed, and an 
overall low burning efficiency was established. NSE elements, but also IMEs, 
are significantly less abundant than in ordinary SNe~Ia; instead, most of the 
ejecta above $6000$ km\,s$^{-1}$ are made of oxygen, and are hence either 
unburned or the result of highly incomplete carbon burning. Furthermore, in the 
$-6$ and $-5$\,d spectra only traces of Fe are found, one order of magnitude 
less than expected for unprocessed material with solar composition. Possible 
implications of this low metallicity on the nature of the progenitor were 
discussed in Section~\ref{Element abundances from synthetic spectra}.

It is presently very difficult to decide whether 91bg-like SNe~Ia form a separate 
group, distinct from ordinary SNe~Ia and possibly descending from different 
progenitors or explosion mechanisms, or whether they are just the extreme end 
of a continuous distribution of objects, with their characteristic appearance 
owing to lower temperatures in conjunction with a different chemical composition, 
both caused by the overall low burning efficiency. 
To tackle this problem will require a new statistical analysis similar to that 
of \citet{Benetti05} once a bigger sample of rapidly-declining SNe~Ia will be 
available, and more detailed spectral modelling of the data sets already published.

\section*{Acknowledgments}

S.T. is grateful to Thomas Matheson for kindly providing access to 
unpublished spectra of SN~1998de obtained with the 1.5\,m Tillinghast 
Telescope + FAST spectrograph.

This work has been supported by the European Union's Human Potential
Programme ``The Physics of Type Ia Supernovae," under contract
HPRN-CT-2002-00303. M.H. acknowledges support from the Centro de 
Astrof\'isica FONDAP 15010003, Proyecto Fondecyt 1060808, and 
N\'ucleo Milenio P06-045-F. G.P. acknowledges support by the 
Proyecto Fondecyt 3070034.

The paper is based on observations collected at the 2.5\,m du Pont and 
the 1.0\,m Swope Telescopes (Las Campanas, Chile), the 2.2\,m Telescope 
of the Centro Astron\'omico Hispano Alem\'an (Calar Alto, Spain), the 
Asiago 1.82\,m and the Loiano 1.52\,m Telescopes (INAF observatories, 
Italy), the Italian 3.58\,m Telescopio Nazionale Galileo and the 2.0\,m 
Liverpool Telescope (La Palma, Spain), the Wendelstein 0.8\,m Telescope 
(Bavaria, Germany), and the ESO Very Large Telescope (Cerro Paranal, Chile). 
The Telescopio Nazionale Galileo is operated by the Fundaci\'on Galileo 
Galilei of the INAF (Instituto Nazionale di Astrofisica) at the Spanish 
Observatorio del Roque de los Muchachos of the Instituto de Astrofisica 
de Canarias.
Our thanks go to the support astronomers at the Telescopio Nazionale
Galileo, the Loiano 1.52\,m Telescope and the 2.2\,m Telescope in Calar 
Alto for performing the follow-up observations of SN~2005bl. 

This research made use of the NASA/IPAC Extragalactic Database (NED)
which is operated by the Jet Propulsion Laboratory, California
Institute of Technology, under contract with the National Aeronautics
and Space Administration, and the Lyon-Meudon Extragalactic Database
(LEDA), supplied by the LEDA team at the Centre de Recherche
Astronomique de Lyon, Observatoire de Lyon.

\addcontentsline{toc}{chapter}{Bibliography}
\markboth{Bibliography}{Bibliography}
\bibliographystyle{mn2e}

\bsp

\label{lastpage}

\end{document}